\documentclass[%
twocolumn,
reprint,
superscriptaddress,
nofootinbib,
amsmath,amssymb,
aps,
]{revtex4-1}

\usepackage{color}
\usepackage{graphicx}
\usepackage[caption=false]{subfig}
\usepackage{dcolumn}
\usepackage{bm}
\usepackage{braket}
\usepackage{fouridx}
\usepackage[
colorlinks=true,
linkcolor=blue,
citecolor=blue,
urlcolor=blue,
]{hyperref}
\usepackage[capitalise]{cleveref}
\usepackage{ulem}

\newcommand{\bes}{\begin{subequations}}
\newcommand{\ees}{\end{subequations}}
\newcommand{\stack}[1]{\begin{gathered} #1\end{gathered}}
\newcommand{\Tr}{\mathop{\mathrm{Tr}} \nolimits}
\newcommand{\sgn}{\mathop{\mathrm{sgn}}}
\newcommand{\Span}{\mathop{\mathrm{span}}}
\newcommand{\rmt}{\mathrm{t}}
\newcommand{\rmb}{\mathrm{b}}
\newcommand{\uar}{\uparrow}
\newcommand{\dar}{\downarrow}
\newcommand{\rar}{\rightarrow}
\newcommand{\lar}{\leftarrow}
\renewcommand{\Re}{\mathop{\mathrm{Re}}}

\begin{document}


\title{Phase transitions in the frustrated Ising ladder with stoquastic and nonstoquastic catalysts} 

\author{Kabuki Takada}
\thanks{Present address: Japan Digital Design, Inc., 3-3-5 Nihonbashi-hongokucho, Chuo, Tokyo 103-0021, Japan}
\email{\\mtcamkxacdiaiki@gmail.com}
\affiliation{Department of Physics, Tokyo Institute of Technology, Yokohama, Kanagawa 226-8503, Japan}

\author{Shigetoshi Sota}
\affiliation{Computational Materials Science Research Team, RIKEN Center for Computational Science (R-CCS), Kobe, Hyogo 650-0047, Japan}

\author{Seiji Yunoki}
\affiliation{Computational Condensed Matter Physics Laboratory, RIKEN Cluster for Pioneering Research (CPR), Wako, Saitama 351-0198, Japan}
\affiliation{Computational Quantum Matter Research Team, RIKEN Center for Emergent Matter Science (CEMS), Wako, Saitama 351-0198, Japan}
\affiliation{Computational Materials Science Research Team, RIKEN Center for Computational Science (R-CCS), Kobe, Hyogo 650-0047, Japan}

\author{\\Bibek Pokharel}
\affiliation{Department of Physics, University of Southern California, Los Angeles, California 90089, USA}
\affiliation{Center for Quantum Information Science \& Technology, University of Southern California, Los Angeles, California 90089, USA}

\author{Hidetoshi Nishimori}
\affiliation{Institute of Innovative Research, Tokyo Institute of Technology, Yokohama, Kanagawa 226-8503, Japan}
\affiliation{Graduate School of Information Sciences, Tohoku University, Sendai, Miyagi 980-8579, Japan}
\affiliation{RIKEN Interdisciplinary Theoretical and Mathematical Sciences Program (iTHEMS), Wako, Saitama 351-0198, Japan}

\author{Daniel A. Lidar}
\affiliation{Department of Physics, University of Southern California, Los Angeles, California 90089, USA}
\affiliation{Center for Quantum Information Science \& Technology, University of Southern California, Los Angeles, California 90089, USA}
\affiliation{Department of Electrical \& Computer Engineering, and Department of Chemistry, University of Southern California, Los Angeles, California 90089, USA}

\date{\today}

\begin{abstract}
The role of nonstoquasticity in the field of quantum annealing and adiabatic quantum computing is an actively debated topic. We study a strongly-frustrated quasi-one-dimensional quantum Ising model on a two-leg ladder to elucidate how a first-order phase transition with a topological origin is affected by interactions of the $\pm XX$-type. Such interactions are sometimes known as stoquastic (negative sign) and nonstoquastic (positive sign) ``catalysts''. Carrying out a symmetry-preserving real-space renormalization group analysis and extensive density-matrix renormalization group computations, we show that the phase diagrams obtained by these two methods are in qualitative agreement with each other and reveal that the first-order quantum phase transition of a topological nature remains stable against the introduction of both $XX$-type catalysts. This is the first study of the effects of nonstoquasticity on a first-order phase transition between topologically distinct phases. Our results indicate that nonstoquastic catalysts are generally insufficient for removing topological obstacles in quantum annealing and adiabatic quantum computing.
\end{abstract}

\pacs{Valid PACS appear here}

\maketitle

\section{Introduction} \label{sec.intro}

Quantum annealing 
exploits quantum-mechanical fluctuations to solve combinatorial optimization problems~\cite{kadowaki1998,Brooke1999,farhi2001,Santoro2002,Santoro:2006,Morita2008,Albash2018,Hauke2020}. In a typical formulation, the combinatorial optimization problem one wants to solve is expressed as an Ising model, represented in terms of the $z$ components of the Pauli matrices, and quantum fluctuations are introduced as a transverse field (sometimes called the $X$ term), the sum of the $x$ components of the Pauli matrices over all sites. One of the bottlenecks of quantum annealing under unitary Schr\"odinger dynamics is a first-order quantum phase transition as a function of the relative weight of the Ising Hamiltonian and the $X$ term, which exists in almost all cases of interest. At a first-order quantum phase transition, the energy gap between the ground state and the first excited state closes exponentially as a function of the system size except for a few very limited cases~\cite{Tsuda2013,Pfeuty1970,Laumann2012}. This leads to exponentially long computation times within the framework of adiabatic time evolution, according to the adiabatic theorem of quantum mechanics~\cite{Jansen:07,mozgunov2020quantum}.\footnote{In a strict sense the adiabatic theorem only provides an upper bound, so in principle it is possible that the computation time scales more favorably, but this needs to be analyzed on a case-by-case basis.} There has been a large amount of effort to circumvent this difficulty including, but not limited to, the introduction of nonstoquastic catalysts~\cite{FarhiAQC:02,Seki2012,Seoane2012,Crosson2014,Seki2015,Hormozi2017,Nishimori2017,Albash2019,Takada2020,Matsuura2020,Crosson2020}, inhomogeneous field driving~\cite{Farhi2011,Dickson2011,Dickson2012,Susa2018,Mohseni2018,Susa2018a,Adame2020}, reverse annealing~\cite{Perdomo-Ortiz2010,Chancellor2017,Ohkuwa2018,Yamashiro2019,King2018,Ottaviani2018,Marshall2019,Venturelli2019,Ikeda2019,Rocutto2020,Passarelli2020,Arai2021},  pausing~\cite{Marshall2019,Passarelli2019,Kadowaki2019,chen2020pausing}, and diabatic quantum annealing~\cite{crosson2020prospects}.
In the present paper, we focus on the approach of nonstoquastic catalysts.

A Hamiltonian is called stoquastic if there is a choice of a local (tensor-product) basis in which the Hamiltonian off-diagonal matrix elements are all real and nonpositive~\cite{Bravyi2008}. 
Otherwise the Hamiltonian is called nonstoquastic, and the inevitable positive or complex off-diagonal matrix elements of the Hamiltonian lead to the infamous sign problem~\cite{Loh:1990aa,Gupta2019}. We note that even when the Hamiltonian is stoquastic, but it is presented in a form in which this stoquasticity is unapparent, the problem of deciding whether there exists a local transformation to a basis that ``cures the sign problem'' (makes it stoquastic) by making all of the Hamiltonian matrix elements real and nonpositive, is NP-complete~\cite{Marvian:2019aa,klassen2019hardness}.

Many interesting quantum models are stoquastic, e.g., the transverse-field Ising model of conventional quantum annealing, the Bose-Hubbard model, and particles subject to a position-dependent potential, which includes superconducting flux
circuits if we associate flux with position and charge with momentum~\cite{Bravyi2008,halverson2020efficient}. The partition-function decomposition of a stoquastic Hamiltonian leads to a sum of nonnegative weights (i.e., the absence of sign problem, which means that the path-integral quantum Monte Carlo algorithm~\cite{Suzuki1976} can typically be used efficiently for sampling tasks). Moreover, the ground state of a stoquastic Hamiltonian has only nonnegative amplitudes by the Perron-Frobenius theorem. These two observations are often cited as reasons that stoquastic Hamiltonians should be considered to be less computationally powerful than their nonstoquastic counterparts. In addition, it is well known that nonstoquastic adiabatic quantum computation is universal~\cite{Aharonov:04,MLM:06,Gosset:2014rp}, while universality in the stoquastic case requires excited states~\cite{Jordan:2010fk}. From a computational complexity perspective, the class StoqMA associated with the task of estimating the ground-state energy of stoquastic local Hamiltonians is known to be no harder than QMA and no easier than MA (and hence no easier than NP)~\cite{Bravyi:2006aa} (see Ref.~\cite{Albash2018} for more background).

However, recently evidence has been mounting that nonstoquasticity does not have a clear-cut computational benefit even in the ground state setting. For example, very recently examples were found for which evolution in the ground state of a stoquastic Hamiltonian can solve a problem superpolynomially~\cite{hastings2020power} or even subexponentially~\cite{Gilyen:2020aa} faster than is possible classically. In addition, there is both theoretical and numerical evidence that adiabatic paths based on nonstoquastic Hamiltonians generically have smaller gaps between the ground state and the first excited state, with the implication that they are less useful than stoquastic Hamiltonians for adiabatic quantum optimization~\cite{Crosson2020}. In this work we contribute further evidence that nonstoquasticity is not necessarily useful for efficiently solving a problem.

The setting for our work is the conventional transverse-field Ising model used in quantum annealing, but with added antiferromagnetic $XX$ interactions ($XX$ terms with positive coefficients). Such terms, which can turn a stoquastic Hamiltonian into a nonstoquastic one, are sometimes called nonstoquastic ``catalysts'' when they are turned on only at intermediate times (typically as $s(1-s)$, where $s\in[0,1]$ is the dimensionless time along the anneal)~\cite{Albash2018,Albash2019}, a terminology inspired by its use in entanglement theory~\cite{Jonathan:1999aa}.\footnote{The first use of a nonstoquastic catalyst in adiabatic quantum optimization dates back to Ref.~\cite{FarhiAQC:02}, which referred to it as an ``extra piece of the Hamiltonian.'' In that work the catalyst did not have the specific $XX$ form but was a random Hermitian (hence nonlocal) matrix.}

One of the outstanding features of the introduction of an $XX$-type nonstoquastic catalyst lies in the universality and QMA-completeness of the resulting Hamiltonian, assuming independent control of each term in the Hamiltonian~\cite{Biamonte2008}. Our focus is, however, on another aspect, namely, that a certain set of nonstoquastic catalysts of the $XX$ type is known to reduce the order of quantum phase transitions from first to second~\cite{Seki2012,Seoane2012,Seki2015,Nishimori2017,Albash2019,Takada2020}. This means that adiabatic evolution converges to the ground state of the final Hamiltonian in quantum annealing with an exponential speedup relative to the stoquastic case. However, it does not guarantee a quantum speedup relative to classical solution methods, and in addition examples are known where $XX$-type nonstoquastic catalysts do not lead to performance improvements~\cite{Hormozi2017,Farhi2011,Takada2020,Crosson2020}. All of these studies concern problems with relatively simple phase transitions without a drastic change in the topological structure of thermodynamic phases.

The frustrated ladder model introduced by Laumann \textit{et~al.}~\cite{Laumann2012} is an exceptional case in that it undergoes a first-order quantum phase transition between topologically different phases despite the simplicity of the problem, which is defined on a quasi-one-dimensional two-leg ladder with nearest-neighbor interactions. The model is closely related to a dimer problem defined on a dual ladder, through which the topological aspect of its phase transition can be understood intuitively, as will be detailed in Sec.~\ref{sec.model}. For this reason, we study the effects of nonstoquastic $XX$ catalysts on the phase transition of the frustrated Ising ladder, and the stoquastic case is also treated for completeness.

We employ theoretical and numerical tools, i.e., the real-space renormalization group (RG) method and the density-matrix renormalization group (DMRG) method, to study the structure of the phase diagram with and without $XX$ catalysts of both signs (stoquastic and nonstoquastic). We find that the phase diagrams obtained by the two methods are qualitatively consistent with each other and conclude that the $XX$-type catalysts, both stoquastic and nonstoquastic, keep the order of the phase transition intact. This is the first example where the role of stoquastic and nonstoquastic catalysts is revealed systematically in a low-dimensional system that exhibits a first-order phase transition with a change of the topological structure.

This paper is organized as follows: In the next section we formulate the problem and describe the known properties of the model system. In Sec.~\ref{sec.rsrg} we analyze the problem via the real-space RG method. The results of extensive numerical computations are explained in Sec.~\ref{sec.dmrg} and are compared with those from the real-space RG method. We conclude in Sec.~\ref{sec.conc}. Additional technical details are given in the Appendices.

\section{Model} \label{sec.model}

We consider the frustrated Ising ladder with transverse fields and $XX$ interactions. This is a generalization of the model proposed and analyzed in Ref.~\cite{Laumann2012}, where the transverse field was uniformly applied to all sites and no $XX$ interactions were taken into account. This section describes the definition of the model and its basic properties in the case without $XX$ interactions, and is largely a recapitulation of Ref.~\cite{Laumann2012}.

\subsection{Definition of the model} \label{ssec.model_defmodel}

\begin{figure}
\includegraphics{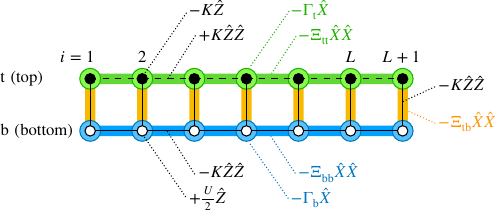}
\caption{Frustrated Ising ladder with transverse fields and $XX$ interactions. Spin-$1/2$ particles are located at the black circles on the top row and white circles on the bottom row. The black solid lines are ferromagnetic interactions of magnitude $K$ and the black dashed lines are antiferromagnetic interactions of magnitude $K$. Local longitudinal fields $K$ and $U/2$ are applied at each black circle and each white circle, respectively, and have opposite directions. A transverse field $\Gamma _\rmt$ is appended at every site on the top row and $\Gamma _\rmb$ on the bottom row. An $XX$ interaction $\Xi _{\rmt\rmt}$ is applied on each horizontal bond on the top row, $\Xi _{\rmb\rmb}$ on each horizontal bond on the bottom row, and $\Xi _{\rmt\rmb}$ on each vertical bond between the two rows. The interactions and fields are indicated by operators in the figure, where the subscripts of the Pauli operators $\hat{X}_{a,i}$, $\hat{Y}_{a,i}$, and $\hat{Z}_{a,i}$ are omitted. Periodic boundary conditions are imposed by identifying the sites at the horizontal position $i=L+1$ with those at $i=1$.} \label{fig.model_frustisinglad}
\end{figure}

As depicted in Fig.~\ref{fig.model_frustisinglad}, the system is composed of qubits (spin-1/2 particles) located on sites of a two-leg ladder. The ladder has a top row ($\rmt$) and a bottom row ($\rmb$). Nearest-neighbor interactions are of magnitude $K>0$ with ferromagnetic (solid lines) or antiferromagnetic (dashed lines) signs. Local longitudinal fields applied to the top and bottom rows have magnitudes $K$ and $U/2>0$, respectively, and are oppositely directed. With transverse fields ($X$ terms) of magnitude $\Gamma _\rmt$ or $\Gamma _\rmb$ and transverse interactions ($XX$ terms) with magnitude $\Xi _{\rmt\rmt}$, $\Xi _{\rmt\rmb}$, or $\Xi _{\rmb\rmb}$, the Hamiltonian is written as
\begin{align}
\hat{H} & =\sum _{i=1}^L \biggl[ K(\hat{Z}_{\rmt ,i} \hat{Z}_{\rmt ,i+1} -\hat{Z}_{\rmb ,i} \hat{Z}_{\rmb ,i+1} -\hat{Z}_{\rmt ,i} \hat{Z}_{\rmb ,i} -\hat{Z}_{\rmt ,i}) \notag \\
& \hphantom{{} =\sum\biggl[} +\frac{U}{2} \hat{Z}_{\rmb ,i} -(\Gamma _\rmt \hat{X}_{\rmt ,i} +\Gamma _\rmb \hat{X}_{\rmb ,i}) \notag \\
& \hphantom{{} =\sum\biggl[} -(\Xi _{\rmt\rmt} \hat{X}_{\rmt ,i} \hat{X}_{\rmt ,i+1} +\Xi _{\rmb\rmb} \hat{X}_{\rmb ,i} \hat{X}_{\rmb ,i+1} \notag \\
& \hphantom{{} =\sum\biggl[ -(} +\Xi _{\rmt\rmb} \hat{X}_{\rmt ,i} \hat{X}_{\rmb ,i})\biggr] , \label{eq.model_hfrustisinglad}
\end{align}
where $\hat{X}_{a,i}$, $\hat{Y}_{a,i}$, and $\hat{Z}_{a,i}$ are the Pauli operators at sites $i=1,\dots ,L$ on row $a=\rmt ,\rmb$. We assume the length $L$ to be even and impose the periodic boundary conditions
\begin{equation}
\hat{X}_{a,L+1} =\hat{X}_{a,1} ,\quad\hat{Y}_{a,L+1} =\hat{Y}_{a,1} ,\quad\hat{Z}_{a,L+1} =\hat{Z}_{a,1} .
\end{equation}
The classical part of $\hat{H}$ (the terms involving $Z$ operators) is highly frustrated due to the competition between positive and negative interactions as well as between interactions and longitudinal fields.

Let us discuss the stoquasticity of the Hamiltonian $\hat{H}$.
As is easily checked, $\hat{H}$ is stoquastic for $\Xi _{\rmt\rmt} , \Xi _{\rmb\rmb} , \Xi _{\rmt\rmb} \geq 0$. Additionally, when $\Xi _{\rmt\rmt} <0$, $\Xi _{\rmb\rmb} <0$, or $\Xi _{\rmt\rmb} <0$, there are cases where a local curing transformation makes the Hamiltonian stoquastic. For example, if $\Gamma _\rmt =\Gamma _\rmb =0$ and $\Xi _{\rmt\rmt} ,\Xi _{\rmb\rmb} ,\Xi _{\rmt\rmb} \leq 0$, consider the transformation obtained by conjugating some qubits by $\hat{Z}_{a,i}$, an operation under which $\hat{X}_{a,i} \mapsto -\hat{X}_{a,i}$. Then, the following is a curing transformation: for odd $i$, conjugate the qubits on the top row and for even $i$, conjugate the bottom row. This transformation is equivalent to flipping the signs of $\Xi _{\rmt\rmt}$, $\Xi _{\rmb\rmb}$, and $\Xi _{\rmt\rmb}$ when $\Gamma _\rmt =\Gamma _\rmb =0$.

It can be shown~\cite{Takada2021} that $\hat{H}$ remains nonstoquastic under single-qubit Clifford transformations if the following set of conditions is satisfied:
\begin{equation}
\left\{
\begin{aligned}
& \Gamma _\rmt ,\Gamma _\rmb >0, \\
& \Xi _{\rmt\rmt} <0\vee\Xi _{\rmb\rmb} <0\vee\Xi _{\rmt\rmb} <0, \\
& \lvert U/2\rvert ,\lvert\Xi _{\rmt\rmb} \rvert <K.
\end{aligned}
\right. 
\label{eq:non-stoq-cond}
\end{equation}
Since the general problem of deciding whether a local curing transformation exists is NP-complete even for single-qubit Clifford transformations~\cite{Marvian:2019aa,klassen2019hardness}, we do not consider here the nonstoquasticity of $\hat{H}$ in more general cases than the conditions given by Eq.~\eqref{eq:non-stoq-cond}. In the following, we refer to $\hat{H}$ as nonstoquastic if there is no curing transformation that is a product of single-qubit Clifford unitaries, and as stoquastic otherwise.

\subsection{Phase diagram for the case without $XX$ terms} \label{ssec.model_phasediagramwoxx}

\begin{figure*}
\includegraphics{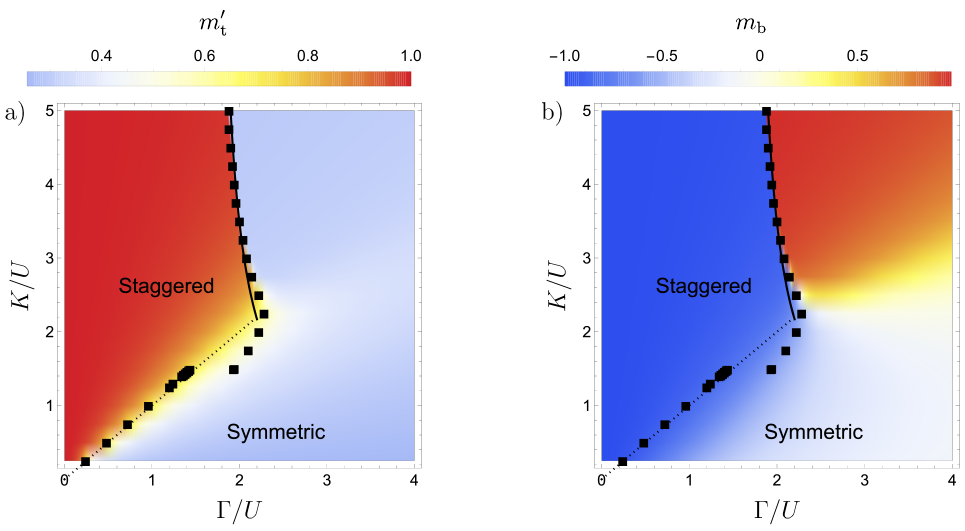}
\caption{Phase diagram of the Ising ladder with uniform transverse field $\Gamma$ and no $XX$ terms, where the staggered magnetization of the top row $m'_\rmt$ is color-coded in (a) and the magnetization of the bottom row $m_\rmb$ in (b). The black squares in (a) and (b) are the locations of the minimum energy gap for fixed values of $K/U$. Here, we set the system size to $L=10$. In both of (a) and (b), the first- and second-order phase boundaries predicted by perturbation theory are shown as solid and dotted lines, respectively. The first-order transition line for $K/U\gg 1$ is $\Gamma /U\approx 1/c+U/(4Kc^3)$ ($c\approx 0.6$) and the second-order transition line for $K/U\ll 1$ is $\Gamma /U\approx K/U$~\cite{Laumann2012}.} \label{fig.phase_diagram_original}
\end{figure*}

Laumann \textit{et~al.}~\cite{Laumann2012} studied the phase diagram in the case of the uniform transverse field $\Gamma _\rmt =\Gamma _\rmb =\Gamma$ and no $XX$ interactions $\Xi _{\rmt\rmt} =\Xi _{\rmb\rmb} =\Xi _{\rmt\rmb} =0$ using numerical diagonalization of small-size systems and perturbation from the large-$K$ and small-$K$ limits. We confirm their findings in Fig.~\ref{fig.phase_diagram_original}, which shows that a first-order transition exists as a function of $\Gamma /U$ with $K/U$ fixed to a large value, while a second-order transition appears for small $K/U$. The values of $\Gamma /U$ that minimize the energy gap between the ground state and the first excited state (which we refer to henceforth as ``minimum gap points'') for $L=10$ and fixed values of $K/U$ are indicated in Fig.~\ref{fig.phase_diagram_original} by black squares. According to perturbation theory, the first-order transition line for $K/U\gg 1$ is $\Gamma /U\approx 1/c+U/(4Kc^3)$ ($c\approx 0.6$) and the second-order transition line for $K/U\ll 1$ is $\Gamma /U\approx K/U$~\cite{Laumann2012}. These two lines meet at $(\Gamma /U,K/U)\approx (2.2,2.2)$.

Note that the perturbation theory prediction agrees with the location of the minimum gap points for $K/U\gtrsim 2.2$ or $K/U\lesssim 1.5$, but the agreement breaks down for $1.5\lesssim K/U\lesssim 2.2$, where the numerically computed locations of the minimum gap deviate from the perturbative second-order transition line. A jump in the minimum gap points is observed at $K/U\approx 1.5$ between $\Gamma /U\approx 1.5$ and $1.9$. We provide an explanation of this phenomenon in terms of the appearance of a ``double-well'' in  the energy gap as a function of $\Gamma/U$ at the critical value $K/U\approx 1.5$, which is the origin of the observed discontinuity. See Appendix~\ref{app:gap} for additional details.

Figure~\ref{fig.phase_diagram_original} also shows the staggered magnetization of the top row
\begin{equation}
m'_\rmt =\Braket{\left\lvert\frac{1}{L} \sum _{i=1}^L (-1)^i \hat{Z}_{\rmt ,i} \right\rvert}
\end{equation}
and the magnetization of the bottom row
\begin{equation}
m_\rmb =\Braket{\frac{1}{L} \sum _{i=1}^L \hat{Z}_{\rmb ,i}}
\end{equation}
for $L=10$, where $\braket{\cdots}$ denotes the expectation value in the ground state. We see that $m'_\rmt$ and $m_\rmb$ have discontinuities as functions of $\Gamma /U$ for large $K/U$, which indicate the existence of the first-order transition. On the other hand, $m'_\rmt$ and $m_\rmb$ change continuously around the second-order transition, which occurs as $\Gamma /U$ is decreased with $K/U$ fixed to a small value.

The ``symmetric'' and ``staggered'' phases shown in Fig.~\ref{fig.phase_diagram_original} are associated with columnar and staggered configurations, which are defined as follows:
\begin{itemize}
\item{Columnar configurations are product states of local eigenstates of $\hat{Z}_{a,i}$ in which all the bottom-row spins are up and there are no nearest-neighbor (consecutive) down spins on the top row.}
\item{Staggered configurations are product states of local eigenstates of $\hat{Z}_{a,i}$ in which all the bottom-row spins are down and nearest-neighbor top-row spins are antiparallel (antiferromagnetically ordered).}
\end{itemize}

When $K$ is large, the two phases can be characterized by perturbation theory in $K^{-1}$. The leading-order part of the state in the symmetric phase is a superposition of columnar configurations (e.g., the red arrows in the top panel of Fig.~\ref{fig.model_mapisingdimer}), while that in the staggered phase has staggered configurations (red arrows in the middle and bottom panels of Fig.~\ref{fig.model_mapisingdimer}). The name ``symmetric phase'' means that all columnar spin configurations have comparable amplitudes with the wave function exhibiting translational invariance, and this invariance is broken in the staggered phase. The system is in the symmetric phase for large $\Gamma /U$ and is in the staggered phase for small $\Gamma /U$. Reference~\cite{Laumann2012} also verified that the energy gaps at the first-order transition points decay exponentially as the length $L$ grows and that the decay rate is proportional to $\ln (K/\Gamma )$. This can be intuitively understood from the fact that the transition from a columnar configuration to a staggered configuration requires flipping all the bottom spins as shown in Fig.~\ref{fig.model_mapisingdimer}.

As reviewed in Appendix~\ref{ssec.rgeqslargek_preanl}, the ground states of the nonperturbative Hamiltonian proportional to $K$,
\begin{equation}
\hat{H}^{(0)} =K\sum _{i=1}^L (\hat{Z}_{\rmt ,i} \hat{Z}_{\rmt ,i+1} -\hat{Z}_{\rmb ,i} \hat{Z}_{\rmb ,i+1} -\hat{Z}_{\rmt ,i} \hat{Z}_{\rmb ,i} -\hat{Z}_{\rmt ,i}),
\end{equation}
are the columnar and staggered configurations. In the presence of the perturbative terms with the coefficients $U$ and $\Gamma$, the degeneracy of $\hat{H}^{(0)}$ is lifted as described above.

The number of columnar configurations is the sum of two Fibonacci numbers $F_{L-1} +F_{L-3}$, which is exponentially large in the length $L$. Here, $F_L$ is defined by the recurrence relation $F_L =F_{L-1} +F_{L-2}$ and the initial values $F_1 =2$ and $F_2 =3$. On the other hand, there are only two staggered configurations. These basic properties are confirmed in Appendix~\ref{ssec.rgeqslargek_preanl}.

In the opposite limit $U\to\infty$, the bottom spins are fixed to down, and the top row of the original Ising ladder~\eqref{eq.model_hfrustisinglad} with $\Gamma _\rmt =\Gamma _\rmb =\Gamma$ and $\Xi _{\rmt\rmt} =\Xi _{\rmb\rmb} =\Xi _{\rmt\rmb} =0$ is effectively an antiferromagnetic Ising chain in a transverse field, $\sum _{i=1}^L (K\hat{Z}_{\rmt ,i} \hat{Z}_{\rmt ,i+1} -\Gamma\hat{X}_{\rmt ,i})$. This results in a second-order transition approximately at $\Gamma =K$ for small $K/U$, as shown in Fig.~\ref{fig.phase_diagram_original}. The symmetric and staggered phases of the ladder correspond to the quantum paramagnetic and antiferromagnetic phases on the top row, respectively.

\subsection{Dimer model on the dual lattice} \label{ssec.model_dimermodel}

\begin{figure}
\includegraphics{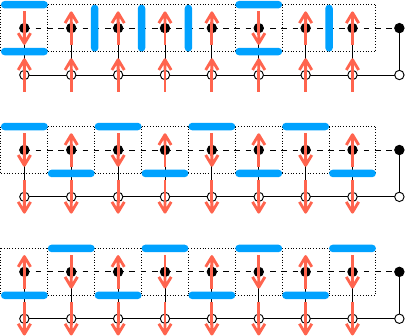}
\caption{Examples of the correspondence between states in the frustrated Ising model and the dimer model on two-leg ladders. The lattice for the former model is depicted in a similar way to Fig.~\ref{fig.model_frustisinglad} (namely, the black circles, the black solid lines, and the black dashed lines are longitudinal fields, ferromagnetic interactions, and antiferromagnetic interactions, respectively, of equal magnitude $K$), while the dual lattice for the latter model is indicated by the dotted lines. The rightmost points are identified with the leftmost points for each lattice, which is subject to periodic boundary conditions. The red arrows are spins and the blue thick line segments are dimers. The topological number is $w=0,+1,-1$ from top to bottom, the first being a columnar configuration and the second and the third being staggered ones.} \label{fig.model_mapisingdimer}
\end{figure}

A significant property of the frustrated Ising ladder is that in the limit of large frustration $K\to\infty$ the model is equivalent to the quantum dimer model on a two-leg ladder, as shown in Ref.~\cite{Laumann2012}. We can find this equivalence by locating dimers on the dual lattice according to the types and positions of frustration exhibited by the columnar and staggered configurations. The dual lattice is a two-leg ladder whose position is shifted horizontally and vertically by half of the unit length from the Ising ladder (see Fig.~\ref{fig.model_mapisingdimer}). Types of frustration in the columnar and staggered configurations are classified as (i) a down spin on the top row (opposite to the longitudinal field), (ii) a horizontally aligned ferromagnetic pair on the top row (opposite to the antiferromagnetic interaction), and (iii) a vertically aligned antiferromagnetic pair (opposite to the ferromagnetic interaction). In the dimer picture these three types become (i) a top horizontal dimer, (ii) a vertical dimer, and (iii) a bottom horizontal dimer, respectively. In this manner the columnar and staggered configurations in the frustrated Ising ladder are mapped one-to-one onto hardcore dimer coverings on the dual two-leg ladder.

Dimer coverings on a two-leg ladder are classified into three topological sectors: the columnar sector $w=0$ and the two staggered sectors $w=\pm 1$, where $w$ denotes the difference in the number of dimers between the top and bottom rows on an arbitrarily given unit square (plaquette). The fact that one cannot transform a dimer covering into another dimer covering with a different $w$ by a series of local movements of dimers allows us to regard $w$ as a topological number.

From this mapping of the columnar and staggered configurations in the frustrated Ising ladder onto dimer coverings, it follows that the Hamiltonian of the frustrated Ising ladder~\eqref{eq.model_hfrustisinglad} in the limit $K\to\infty$ is equivalent to the following Hamiltonian of the quantum dimer model:
\begin{align}
\hat{H}_\text{dimer} & =U\left(\sum _\vert \Ket{\phantom{\Bigl\lvert} \vert\phantom{\Bigr\rvert}} \Bra{\phantom{\Bigl\lvert} \vert\phantom{\Bigr\rvert}} +2\sum _\square \Ket{\vphantom{\Bigl\lvert} =\vphantom{\Bigr\rvert}} \Bra{\vphantom{\Bigl\lvert} =\vphantom{\Bigr\rvert}} \right) \notag \\
& \hphantom{{} = {}} -\Gamma _\rmt \sum _\square \left(\Ket{\phantom{\Bigl\lvert} \Vert\phantom{\Bigr\rvert}} \Bra{\vphantom{\Bigl\lvert} =\vphantom{\Bigr\rvert}} +\Ket{\vphantom{\Bigl\lvert} =\vphantom{\Bigr\rvert}} \Bra{\phantom{\Bigl\lvert} \Vert\phantom{\Bigr\rvert}} \right) \label{eq.model_hdimer} \\
& \hphantom{{} = {}} -\Xi _{\rmt\rmt} \sum _{\square\!\square} \left(\Ket{\phantom{\Bigl\lvert} =\vert\phantom{\Bigr\rvert}} \Bra{\phantom{\Bigl\lvert} \vert =\phantom{\Bigr\rvert}} +\Ket{\phantom{\Bigl\lvert} \vert =\phantom{\Bigr\rvert}} \Bra{\phantom{\Bigl\lvert} =\vert\phantom{\Bigr\rvert}} \right) , \notag
\end{align}
where a constant energy shift was ignored. The summations $\sum _\vert$, $\sum _\square$, and $\sum _{\square\!\square}$ are performed over vertical lines $\vert$, plaquettes $\square$, and pairs of two neighboring plaquettes $\square\!\square$ on the dual lattice, respectively. The line segments in the kets and bras denote dimers that are located in the summation ``variables'' $\vert$, $\square$, and $\square\!\square$. We call the limit $K\to\infty$ the dimer limit. Note that the terms with the coefficients $\Gamma _\rmb$, $\Xi _{\rmb\rmb}$, and $\Xi _{\rmt\rmb}$ do not appear in the dimer model, because the bottom spins should exhibit complete ferromagnetic order in the limit $K\to\infty$.

Since the only nonvanishing matrix elements of $\hat{H}_\text{dimer}$ are those between states in the columnar sector, it naturally follows that there is a strict (not avoided) energy-level crossing between the columnar and staggered sectors in the quantum dimer model on a two-leg ladder. Indeed, the Hamiltonian $\hat{H}_\text{dimer}$ has a strict level crossing at $\Gamma _\rmt /U\approx 1/0.6$ in the $\Xi _{\rmt\rmt} =0$ case (vanishing $XX$ interactions on the top row) according to numerical diagonalization results for small-size systems~\cite{Laumann2012}. Although for the frustrated Ising ladder with large but finite $K$ the level crossing turns into an avoided crossing [due to nonvanishing transition probabilities to defect states with energy penalties $\mathcal{O}(K)$], the numerical consequence that the energy gap at the first-order transition decays exponentially at least for $\Xi _{\rmt\rmt} =0$ reflects the topological nature relating to the quantum dimer model~\cite{Laumann2012}.

\section{Real-space RG analysis} \label{sec.rsrg}

We perform the real-space RG analysis of the frustrated Ising ladder in the limit of large frustration $K\to\infty$, namely the dimer limit, in Sec.~\ref{ssec.rsrg_largek}. Subsequently we analyze the limit of small frustration $U\to\infty$, in Sec.~\ref{ssec.rsrg_largeu}.

\subsection{Large frustration limit} \label{ssec.rsrg_largek}

We analyze the zero-temperature phase transition of the frustrated Ising ladder in the dimer limit $K\gg U,\Gamma _a ,\Xi _{aa'}$ using the real-space RG method. Technical details are delegated to Appendix~\ref{sec.rgeqslargek}.

Although the standard real-space RG transformation amounts to separating the Hamiltonian into intrablock and interblock terms and projecting the Hilbert space onto a low-energy space of the intrablock Hamiltonian~\cite{Nishimori2011}, this procedure will fail to find a low-energy subspace due to the presence of the strong interblock interactions with magnitude $K$. To circumvent this problem, we employ a real-space RG method in which the projector onto a low-energy space is variationally determined such that the dimer structure is preserved at each RG step. Here, the dimer structure means that the columnar and staggered configurations remain the ground states of the leading-order Hamiltonian in $K^{-1}$. Since the columnar and staggered configurations will be the low-energy states with a small energy splitting near the transition, such a variational ansatz may enable us to extract critical properties of the entire system at zero temperature.

\subsubsection{Effective Hamiltonian}

To write down the RG equations, let us define the generalized Hamiltonian that appears in our RG analysis:
\begin{align}
\hat{H} & =\sum _{\sigma :\text{nondimer}} K_\sigma \ket{\sigma} \bra{\sigma} \notag \\
& \hphantom{{} = {}} +\sum _{i=1}^L \left[\frac{U}{2} \hat{Z}_{\rmb ,i} -\Gamma\hat{X}_{\rmt ,i} +V\hat{Z}_{\rmt ,i} -\Xi\hat{X}_{\rmt ,i} \hat{X}_{\rmt ,i+1} \right] +\hat{O} , \label{eq.rsrg_genhlargek}
\end{align}
where nondimer configurations $\ket{\sigma}$ indicate product states of local eigenstates of $\hat{Z}_{a,i}$ that are neither columnar nor staggered configurations. The coefficients $0<K_\sigma =\mathcal{O} (K)$ stand for energy penalties on nondimer configurations $\sigma$. We denoted $\Gamma _\rmt =\Gamma$ and $\Xi _{\rmt\rmt} =\Xi$ for notational simplicity. The longitudinal field on the top row $V$ is produced after a step of the RG transformation. The operator $\hat{O}$ has a magnitude comparable to $U$, $\Gamma$, $V$, and $\Xi$, but is irrelevant in the sense of RG theory. The couplings $\Gamma _\rmb$, $\Xi _{\rmb\rmb}$, and $\Xi _{\rmt\rmb}$ are included in $\hat{O}$. For a more detailed version of the generalized Hamiltonian, see Eq.~\eqref{eq.rgeqslargek_hall} (although the operators with the coefficients $\Gamma$, $V$, and $\Xi$ are slightly different from Eq.~\eqref{eq.rsrg_genhlargek}, the differences are irrelevant). The bare Hamiltonian~\eqref{eq.model_hfrustisinglad} can be obtained by setting $V=0$ except for a constant energy difference proportional to $K$, as described in Appendix~\ref{ssec.rgeqslargek_preanl}.

\subsubsection{RG equations}

Let us perform the RG transformation repeatedly. We denote the coupling constants that have been renormalized $l$ times by $U(l)$, $\Gamma (l)$, $V(l)$, and $\Xi (l)$. The bare couplings are $U(0)$, $\Gamma (0)$, $V(0)=0$, and $\Xi (0)$. Our RG transformation has the scaling factor $b=3$ (note that the scaling factor $b$ should be an odd number to keep the antiferromagnetic order in the staggered phase through the RG transformation) and the system size scales as $L(l)=b^{-l} L(0)$, where $L(0)$ is the length of the original ladder. As derived in Appendix~\ref{sec.rgeqslargek}, the renormalized coupling constants are determined by the RG equations
\bes
\label{eq.rsrg_rglargek}
\begin{align}
U(l+1) & =3U(l)-\Gamma (l)[2\beta _2 (l)(\beta _1 (l)+z(l)\alpha _1 (l)) \notag \\
& \hphantom{{} =3U(l)-\Gamma (l)[} +(1-z(l)^2)\alpha _1 (l)\alpha _2 (l)] \notag \\
& \hphantom{{} = {}} +V(l)[2-(1-z(l)^2)\alpha _1 (l)^2 -\beta _1 (l)^2] \notag \\
& \hphantom{{} = {}} +2\Xi (l)\beta _2 (l)z(l)\alpha _2 (l), \label{eq.rsrg_rgulargek} \\
\Gamma (l+1) & =\Gamma (l)[2\beta _2 (l)\alpha _2 (l)+z(l)(\alpha _1 (l)^2 -\alpha _2 (l)^2)] \notag \\
& \hphantom{{} = {}} -2V(l)z(l)\alpha _1 (l)\alpha _2 (l)+2\Xi (l)\beta _2 (l)\alpha _1 (l), \label{eq.rsrg_rgglargek} \\
V(l+1) & =\Gamma (l)[2\beta _2 (l)(\beta _1 (l)+z(l)\alpha _1 (l)) \notag \\
& \hphantom{{} =\Gamma (l)[} -(1+z(l)^2)\alpha _1 (l)\alpha _2 (l)] \notag \\
& \hphantom{{} = {}} +V(l)[1-(1+z(l)^2)\alpha _1 (l)^2 +\beta _1 (l)^2] \notag \\
& \hphantom{{} = {}} -2\Xi (l)\beta _2 (l)z(l)\alpha _2 (l), \label{eq.rsrg_rgvlargek} \\
\Xi (l+1) & =\Xi (l)\alpha _2 (l)^2 \beta _2 (l)^2 . \label{eq.rsrg_rgxlargek}
\end{align}
\ees
Here, $\alpha _1 (l),\alpha _2 (l),z(l),\beta _1 (l),\beta _2 (l)\in\mathbb{R}$ are the variational parameters that minimize the function
\begin{align}
& f_{\Gamma (l),V(l),\Xi (l)} (\alpha _1 ,\alpha _2 ,z,\beta _1 ,\beta _2) \notag \\
& =-\Gamma (l)[(\varphi ^2 -z^2)\alpha _1 \alpha _2 +2\beta _2 (\beta _1 +z\alpha _1)] \notag \\
& \hphantom{{} = {}} -V(l)[(\varphi ^2 -z^2)\alpha _1^2 +\beta _1^2]+2\Xi (l)\beta _2 z\alpha _2 \label{eq.rsrg_varfunlargek}
\end{align}
under the constraint
\begin{equation}
\alpha _1^2 +\alpha _2^2 =z^2 +\beta _1^2 +2\beta _2^2 =1, \label{eq.rsrg_normvprlargek}
\end{equation}
where $\varphi =(1+\sqrt{5})/2$ is the golden ratio. This minimization is equivalent to minimizing the trace of the renormalized Hamiltonian in the subspace spanned by the columnar and staggered configurations at each RG step.

Note that the function~\eqref{eq.rsrg_varfunlargek} and the constraint~\eqref{eq.rsrg_normvprlargek} are invariant under the transformations $(\alpha _1 ,\alpha _2 ,z)\mapsto (-\alpha _1 ,-\alpha _2 ,-z)$ and $(z,\beta _1 ,\beta _2)\mapsto (-z,-\beta _1 ,-\beta _2)$. Each of these transformations is equivalent to changing the phase of a variational state included in the projector onto the coarse-grained space and thus leaves the projector invariant. Therefore, the optimal set of the variational parameters $\alpha _1 (l),\alpha _2 (l),z(l),\beta _1 (l),\beta _2 (l)\in\mathbb{R}$ is four-fold degenerate and we choose one of the solutions. The choice among the four solutions does not affect the critical properties of the system since the multiplication of $(\alpha _1 (l),\alpha _2 (l),z(l))$ or $(z(l),\beta _1 (l),\beta _2 (l))$ by $-1$ only flips the sign of the renormalized transverse field $\Gamma (l+1)$, which corresponds to the gauge transformation $(\hat{X}_{a,i} ,\hat{Y}_{a,i} ,\hat{Z}_{a,i})\mapsto (-\hat{X}_{a,i} ,-\hat{Y}_{a,i} ,\hat{Z}_{a,i})$ in the renormalized Hamiltonian.

Let us describe important features of the RG equations. Since Eqs.~\eqref{eq.rsrg_rgglargek}--\eqref{eq.rsrg_rgxlargek} do not contain $U(l)$ explicitly and the variational parameters minimizing $f_{\Gamma (l),V(l),\Xi (l)}$ subject to the constraint~\eqref{eq.rsrg_normvprlargek} can be regarded as functions of $\Gamma (l)$, $V(l)$, and $\Xi (l)$, the renormalized coupling constants $\Gamma (l+1)$, $V(l+1)$, and $\Xi (l+1)$ are expressed as functions of $\Gamma (l)$, $V(l)$, and $\Xi (l)$, not including $U(l)$. In addition, it is convenient to write the coupling constant $U(l)$ as
\begin{equation}
U(l)=3^l (U(0)-\overline{U} (l)),
\end{equation}
which yields the initial value $\overline{U} (0)=0$ and the recurrence relation
\begin{align}
& \overline{U} (l+1) \notag \\
& =\overline{U} (l)+3^{-l-1} \{\Gamma (l)[2\beta _2 (l)(\beta _1 (l)+z(l)\alpha _1 (l)) \notag \\
& \hphantom{{} =\overline{U} (l)+3^{-l-1} \{\Gamma (l)[} +(1-z(l)^2)\alpha _1 (l)\alpha _2 (l)] \notag \\
& \hphantom{{} =\overline{U} (l)+3^{-l-1} \{} -V(l)[2-(1-z(l)^2)\alpha _1 (l)^2 -\beta _1 (l)^2] \notag \\
& \hphantom{{} =\overline{U} (l)+3^{-l-1} \{} -2\Xi (l)\beta _2 (l)z(l)\alpha _2 (l)\} . \label{eq.rsrg_rgblargek}
\end{align}
Now Eqs.~\eqref{eq.rsrg_rgglargek}--\eqref{eq.rsrg_rgxlargek} and \eqref{eq.rsrg_rgblargek} are independent of $U(l)$ as well as $U(0)$. Since $\overline{U} (0)=V(0)=0$, it follows by mathematical induction that $\overline{U} (l)$, $\Gamma (l)$, $V(l)$, and $\Xi (l)$ are determined only by the ``time'' $l$ and the initial values $\Gamma (0)$ and $\Xi (0)$.

Although the penalty constants $K_\sigma$ for nondimer configurations $\sigma$ are also renormalized, their specific values are unimportant for our analysis of the zero-temperature phase transition in the dimer limit. The only essential point is that all the penalty constants are positive and proportional to $K$. This means that any spin configuration that is not mapped onto a dimer covering does not contribute to the leading-order part of the ground state.

\subsubsection{RG flow and fixed points}

\begin{figure*}
\includegraphics{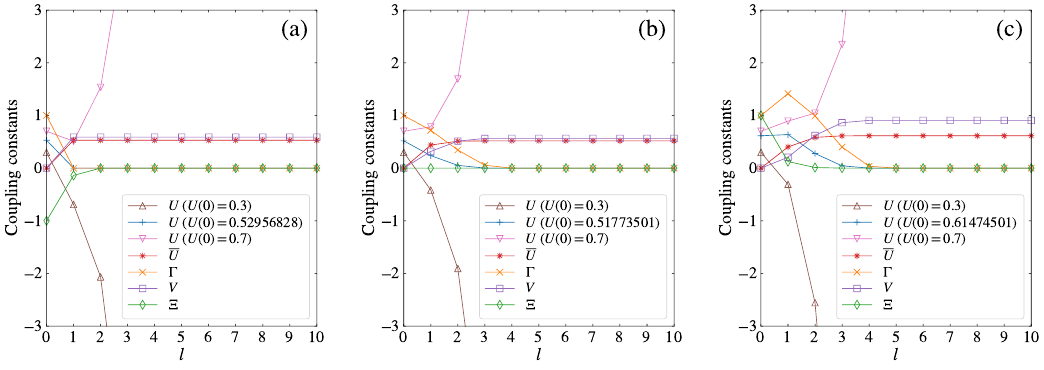}
\caption{Coupling constants $U(l)$, $\overline{U} (l)$, $\Gamma (l)$, $V(l)$, and $\Xi (l)$ as functions of $l$ for the bare couplings $(\Gamma (0),V(0),\Xi (0))=(1,0,-1)$ in (a), $(1,0,0)$ in (b), and $(1,0,1)$ in (c). In each of the cases, $U(0)=0.3,\overline{U}_{\Gamma (0),\Xi (0)} ,0.7$. Note that $\overline{U} (l)$, $\Gamma (l)$, $V(l)$, and $\Xi (l)$ do not depend on $U(0)$.} \label{fig.rsrg_runcpllargek}
\end{figure*}

We show the coupling constants $\overline{U} (l)$, $\Gamma (l)$, $V(l)$, and $\Xi (l)$ as functions of $l$ for the initial values $(\Gamma (0),\Xi (0))=(1,-1),(1,0),(1,1)$ in Fig.~\ref{fig.rsrg_runcpllargek} [see the next paragraph on $U(l)$ and its initial value $U(0)$]. For the behavior of the coupling constants that have other initial values and the behavior of the variational parameters, see Figs.~\ref{fig.rgeqslargek_runcplprmx0}--\ref{fig.rgeqslargek_runcplprmxp} in Appendix~\ref{ssec.rgeqslargek_rgeq}. Note that multiplying all the bare couplings by a positive constant $c$ leaves the variational parameters at each $l$ unchanged and multiplies every renormalized coupling by $c$. This operation corresponds to a rescaling of the vertical axis of each graph in Fig.~\ref{fig.rsrg_runcpllargek}.

We find that $\overline{U} (l)$ and $V(l)$ converge to finite positive values and $\Gamma (l)$ and $\Xi (l)$ vanish in the limit $l\to\infty$. The coupling constant $U(l)$ asymptotically behaves as
\begin{equation}
U(l)\approx 3^l (U(0)-\overline{U}_{\Gamma (0),\Xi (0)})\quad (l\to\infty ), \label{eq.rsrg_aspulargek}
\end{equation}
where $\overline{U}_{\Gamma (0),\Xi (0)} :=\lim _{l\to\infty} \overline{U} (l)$ is the limiting value of $\overline{U} (l)$ determined by the bare couplings $\Gamma (0)$ and $\Xi (0)$ [recall that $V(0)=0$]. Figure~\ref{fig.rsrg_runcpllargek} shows the behavior of $U(l)$ for several values of $U(0)$ including $\overline{U}_{\Gamma (0),\Xi (0)}$. It turns out that the limit of $U(l)$ is
\begin{equation}
U(l)\to\left\{
\begin{alignedat}{2}
& {-\infty} ,\quad && U(0)<\overline{U}_{\Gamma (0),\Xi (0)} , \\
& 0,\quad && U(0)=\overline{U}_{\Gamma (0),\Xi (0)} , \\
& {+\infty} ,\quad && U(0)>\overline{U}_{\Gamma (0),\Xi (0)}
\end{alignedat}
\right.\quad (l\to\infty ). \label{eq.rsrg_limulargek}
\end{equation}
This indicates that the fixed points of the present RG transformation are $(U,\Gamma ,V,\Xi )=(0,0,V,0),(\pm\infty ,0,V,0)$ with $V$ being an arbitrary positive value. Indeed, we can confirm that these points are fixed points by finding that $f_{\Gamma =0,V,\Xi =0} (\alpha _1 ,\alpha _2 ,z,\beta _1 ,\beta _2)$ is minimized at $\alpha _1^2 =\beta _1^2 =1$ and $\alpha _2 =z=\beta _2 =0$ under the constraint~\eqref{eq.rsrg_normvprlargek} and substituting these values into the RG equations~\eqref{eq.rsrg_rglargek}.

\begin{figure*}
\includegraphics{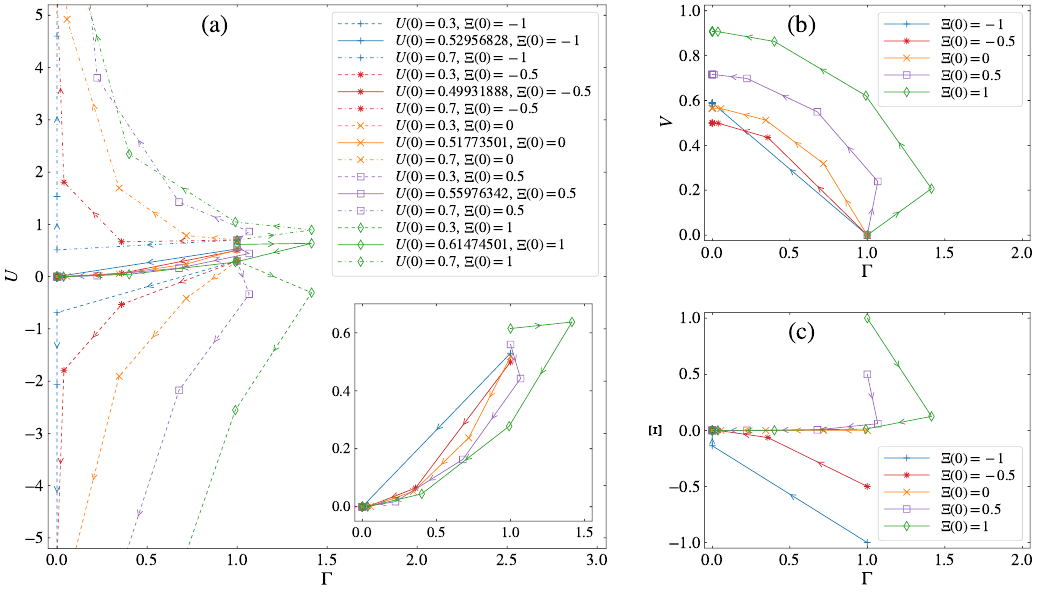}
\caption{RG flow diagrams of the frustrated Ising ladder in the dimer limit $K\to\infty$. We plot $(\Gamma (l),U(l))$, $(\Gamma (l),V(l))$, and $(\Gamma (l),\Xi (l))$ ($l=0,1,\dots ,10$) in (a), (b), and (c), respectively. The bare couplings are set to $\Gamma (0)=1$, $V(0)=0$, $\Xi (0)=0,\pm 0.5,\pm 1$, and $U(0)=0.3,\overline{U}_{\Gamma (0),\Xi (0)} ,0.7$. Note that $\Gamma (l)$, $V(l)$, and $\Xi (l)$ do not depend on $U(0)$. The inset in (a) is a magnification of $(\Gamma (l),U(l))$ for $U(0)=\overline{U}_{\Gamma (0),\Xi (0)}$. The arrows on each line indicate the direction of the RG flow. It can be seen that $l=10$ is sufficient for convergence of $U(l)$ [$U(0)=\overline{U}_{\Gamma (0),\Xi (0)}$], $\Gamma (l)$, $V(l)$, and $\Xi (l)$.} \label{fig.rsrg_rgflowlargek}
\end{figure*}

To gain a clearer understanding of the behavior of the coupling constants, we depict the RG flow diagrams in Fig.~\ref{fig.rsrg_rgflowlargek}. We see that the coupling constants approach the following values in the limit $l\to\infty$:
\begin{align}
& \lim _{l\to\infty} (U(l),\Gamma (l),V(l),\Xi (l)) \notag \\
& =\left\{
\begin{alignedat}{2}
& \left( -\infty ,0,\lim _{l\to\infty} V(l),0\right) ,\quad && U(0)<\overline{U}_{\Gamma (0),\Xi (0)} , \\
& \left( 0,0,\lim _{l\to\infty} V(l),0\right) ,\quad && U(0)=\overline{U}_{\Gamma (0),\Xi (0)} , \\
& \left( +\infty ,0,\lim _{l\to\infty} V(l),0\right) ,\quad && U(0)>\overline{U}_{\Gamma (0),\Xi (0)} ,
\end{alignedat}
\right. \label{eq.rsrg_limcpllargek}
\end{align}
where $\lim _{l\to\infty} V(l)$ is a finite positive value.

The fixed points $(U,\Gamma ,V,\Xi )=(\pm\infty ,0,V,0)$ ($V>0$) are stable (i.e., any coupling constant is irrelevant around these fixed points\footnote{In a strict sense, the coupling constant $V$ is marginal around the fixed points $(U,\Gamma ,V,\Xi )=(0,0,V,0),(\pm\infty ,0,V,0)$ because these are fixed points for any $V>0$. The arbitrariness of $V$ is due to the fact that all the renormalized couplings are multiplied by $c$ under multiplication of all the bare couplings by $c>0$.\label{ftn.rsrg_vmarginal}}) and these correspond to the two phases:
\begin{itemize}
\item{$(U,\Gamma ,V,\Xi )=(-\infty ,0,V,0)$: symmetric phase}
\item{$(U,\Gamma ,V,\Xi )=(+\infty ,0,V,0)$: staggered phase}
\end{itemize}
On the other hand, the fixed point $(U,\Gamma ,V,\Xi )=(0,0,V,0)$ for each $V>0$ is unstable to variations in $U$. Around this fixed point, $U$ is relevant and the other coupling constants are irrelevant.${}^{\ref{ftn.rsrg_vmarginal}}$ Recalling that the scaling factor of our RG transformation is $b=3$ and $U(l)$ is asymptotically proportional to $3^l$ as shown in Eq.~\eqref{eq.rsrg_aspulargek}, we find that the scaling dimension of $U$ around the unstable fixed point $(U,\Gamma ,V,\Xi )=(0,0,V,0)$ is $y_U =1$. Here, the scaling dimension $y_U$ around a fixed point $U=U^*$ is defined as $U(l+1)-U^* \approx b^{y_U} (U(l)-U^*)$ in the vicinity of $U(l)=U^*$ for any scaling factor $b$.

In addition, Eq.~\eqref{eq.rsrg_limcpllargek} shows the bare coupling $U(0)=\overline{U}_{\Gamma (0),\Xi (0)}$ to be the critical point for each $\Gamma (0)$ and $\Xi (0)$, meaning that only the fine-tuned $U(0)=\overline{U}_{\Gamma (0),\Xi (0)}$ flows into the unstable fixed point $(U,\Gamma ,V,\Xi )=(0,0,V,0)$ in the sense of RG theory. Other values of the bare coupling $U(0)$ are absorbed into the stable fixed points [namely, $U(0)<\overline{U}_{\Gamma (0),\Xi (0)}$ into the symmetric phase $(U,\Gamma ,V,\Xi )=(-\infty ,0,V,0)$ and $U(0)>\overline{U}_{\Gamma (0),\Xi (0)}$ into the staggered phase $(U,\Gamma ,V,\Xi )=(+\infty ,0,V,0)$] under repeated application of the RG transformation.

\subsubsection{Phase diagram}

\begin{figure}
\includegraphics{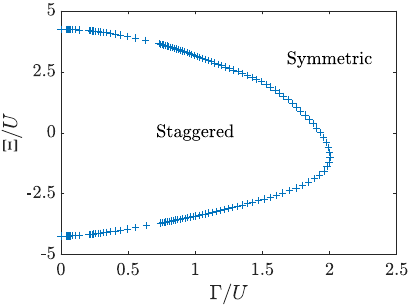}
\caption{Phase diagram of the frustrated Ising ladder in the dimer limit $K\to\infty$ predicted by the real-space RG method. The first-order transition points (blue plus signs) indicate that adding $XX$ catalysts (stoquastic or nonstoquastic) does not remove the first-order transition.} \label{fig.rsrg_phasediagramlargek}
\end{figure}

We now obtain the phase diagram of the frustrated Ising ladder in the dimer limit. In the following, we denote the bare couplings by $U$, $\Gamma$, and $\Xi$ [not $U(0)$, $\Gamma (0)$, and $\Xi (0)$] for notational simplicity. Although there are three couplings $U$, $\Gamma$, and $\Xi$, it suffices to plot the phase diagram in the $\Xi /U$-$\Gamma /U$ plane because dividing these couplings by $U>0$ does not change the phase [note, as already remarked above, that $\overline{U}_{c\Gamma ,c\Xi} =c\overline{U}_{\Gamma ,\Xi}$ for any $c>0$ and thus $\lim _{l\to\infty} U(l)$ is unchanged under multiplication of every bare coupling by $c$]. We can draw the phase diagram by marking the critical points $(\Gamma /U,\Xi /U)=(\Gamma /\overline{U}_{\Gamma ,\Xi} ,\Xi /\overline{U}_{\Gamma ,\Xi})$ for a number of sets of $\Gamma$ and $\Xi$. The resulting phase diagram is shown in Fig.~\ref{fig.rsrg_phasediagramlargek}.

We can deduce from scaling theory~\cite{Nishimori2011}, which is closely related to RG theory, that the phase boundary shown in Fig.~\ref{fig.rsrg_phasediagramlargek} is of first order. The fact that the scaling dimension of the longitudinal field on the bottom row $y_U$ is equal to the spatial dimensionality of the system $d=1$ indicates that the phase transition is of first order, because the correlation function on the bottom row at the critical point does not decay:
\begin{equation}
G_{\rmb i,\rmb j} :=\braket{\hat{Z}_{\rmb ,i} \hat{Z}_{\rmb ,j}}_{U=\overline{U}_{\Gamma ,\Xi}} \sim\lvert i-j\rvert ^{-2(d-y_U)} =1, \label{eq.rsrg_corrlargek}
\end{equation}
where $\braket{\cdots}_{U=\overline{U}_{\Gamma ,\Xi}}$ denotes the expectation value at the critical point $U=\overline{U}_{\Gamma ,\Xi}$ and the approximation is a consequence of scaling theory~\cite{Nishimori2011}. This behavior of the correlation function leads to the value of the anomalous dimension $\eta$, one of the critical exponents defined by $G_{\rmb i,\rmb j} \sim\lvert i-j\rvert ^{2-d-\eta}$:
\begin{equation}
\eta =2+d-2y_U =2-d=1.
\end{equation}

Recall that in general the scaling law for a quantum system is $G_{\rmb i,\rmb j} \sim\lvert i-j\rvert ^{-2(d+z-y_U)}$ with $z$ being the dynamic critical exponent~\cite{Sachdev2011} (which should not be confused with the variational parameter $z$) instead of Eq.~\eqref{eq.rsrg_corrlargek}. However, we find that $z=0$ for the present system, because under a sufficiently large number of RG transformations the system becomes (almost) classical, by elimination of the transverse field $\Gamma$ and the $XX$ interaction $\Xi$.

Consequently, Fig.~\ref{fig.rsrg_phasediagramlargek} demonstrates that there is a first-order phase boundary that completely separates the staggered phase from the symmetric phase on the $\Xi /U$-$\Gamma /U$ plane. In other words, \textit{the first-order phase transition encountered during quantum annealing cannot be removed by stoquastic or nonstoquastic $XX$ catalysts}.

Note that whether the bare Hamiltonian~\eqref{eq.model_hfrustisinglad} is nonstoquastic depends not only on $\Xi$ ($=\Xi _{\rmt\rmt}$) but also on $\Gamma$ ($=\Gamma _\rmt$), $\Gamma _\rmb$, $\Xi _{\rmb\rmb}$, and $\Xi _{\rmt\rmb}$ [recall that the Hamiltonian is nonstoquastic if the fields and interactions satisfy Eq.~\eqref{eq:non-stoq-cond}, whose third condition $\lvert U/2\rvert ,\lvert\Xi _{\rmt\rmb} \rvert <K$ holds in the dimer limit $K\gg U, \Gamma _a ,\Xi _{aa'}$]. For example, consider the case of $\sgn\Gamma _\rmb =\sgn\Gamma$ and $\sgn\Xi _{\rmb\rmb} ,\sgn\Xi _{\rmt\rmb} \in\{ 0,\sgn\Xi\}$, where the sign function is defined as $\sgn x=x/\lvert x\rvert$ for $x\not= 0$ and $\sgn x=0$ for $x=0$. Then, the Hamiltonian is stoquastic if $\Gamma =0\vee\Xi\geq 0$ and nonstoquastic if $\Gamma >0\wedge\Xi <0$. Stoquasticity for $\Gamma =0\wedge\Xi <0$ follows from the curing transformation that flips the signs of $\Xi _{\rmt\rmt}$, $\Xi _{\rmb\rmb}$, and $\Xi _{\rmt\rmb}$, as pointed out in Sec.~\ref{ssec.model_defmodel}. The existence of this transformation indicates that the critical points on $\Gamma =0$, $\overline{U}_{\Gamma =0,\Xi}$, are invariant under a change of the sign of $\Xi$.

It is noteworthy that the transition point derived from our real-space RG method in the absence of $XX$ interactions, $\Gamma /U=\Gamma /\overline{U}_{\Gamma ,\Xi =0} =1.9314900$, is not far from that obtained by numerical diagonalization of the quantum dimer model (whose Hamiltonian is given by Eq.~\eqref{eq.model_hdimer} with $\Xi _{\rmt\rmt} =0$), $\Gamma /U\approx 1/0.6=1.66\dotsm$~\cite{Laumann2012}, in spite of the fact that our RG analysis is not an exact method.

\subsection{Small frustration limit of the Ising chain} \label{ssec.rsrg_largeu}

In this section we step back from the full Ising ladder and analyze the phase transition in the limit of small frustration $U\gg K,\Gamma _a ,\Xi _{aa'}$ at zero temperature. Since the bottom spins are fixed to down in this limit, we obtain an antiferromagnetic Ising chain with a transverse field and $XX$ interactions as an effective model:
\begin{equation}
\hat{H}_\text{chain} =\sum _{i=1}^L (K\hat{Z}_{\rmt ,i} \hat{Z}_{\rmt ,i+1} -\Gamma _\rmt \hat{X}_{\rmt ,i} -\Xi _{\rmt\rmt} \hat{X}_{\rmt ,i} \hat{X}_{\rmt ,i+1}). \label{eq.model_hchain}
\end{equation}
Our purpose in this section is to reveal how the $XX$ interactions affect the second-order transition that appears in the case of no $XX$ interaction and large $U$ (see Fig.~\ref{fig.phase_diagram_original} or Ref.~\cite{Laumann2012}). Readers who are interested only in the effects of the $XX$ interactions on the first-order transition can proceed to the next section on the DMRG calculations.

Similar analyses were conducted by Langari~\cite{Langari1998,Langari2004}. He carried out the real-space RG analysis of the $XXZ$ chain in a magnetic field~\cite{Langari1998}, whose Hamiltonian is equivalent to Eq.~\eqref{eq.model_hchain} with $\sum _i K\hat{Y}_{\rmt ,i} ,\hat{Y}_{\rmt ,i+1}$ added, and derived the zero-temperature phase diagram. In addition, he obtained the zero-temperature phase diagram of the model given by the Hamiltonian~\eqref{eq.model_hchain} for $\Xi _{\rmt\rmt} \leq 0$ using the real-space RG method~\cite{Langari2004}.

To compare the effects of ferromagnetic and antiferromagnetic $XX$ interactions, we perform the real-space RG analysis for positive and negative $\Xi _{\rmt\rmt}$. Instead of the Hamiltonian~\eqref{eq.model_hchain}, we consider the ferromagnetic Ising chain with a transverse field and $XX$ interactions,
\begin{equation}
\hat{H} =\sum _{i=1}^L (-K\hat{Z}_i \hat{Z}_{i+1} -\Gamma\hat{X}_i -\Xi\hat{X}_i \hat{X}_{i+1}), \label{eq.rsrg_hchainferrolargeu}
\end{equation}
which is obtained by the gauge transformation $(\hat{X}_i ,\hat{Y}_i ,\hat{Z}_i)\mapsto (\hat{X}_i ,-\hat{Y}_i ,-\hat{Z}_i)$ for every odd $i$ (we omitted the subscript $\rmt$ in $\Gamma _\rmt$, $\Xi _{\rmt\rmt}$, and the Pauli operators for notational simplicity).

The real-space RG analysis of the model~\eqref{eq.rsrg_hchainferrolargeu} proceeds in the standard manner~\cite{Nishimori2011}, which is detailed in Appendix~\ref{sec.rgeqslargeu}. We first separate the Hamiltonian into intrablock and interblock Hamiltonians after partitioning the chain every two sites, which is valid when $\Xi$ is not a negative large value (if $\Xi <0$ and $\lvert\Xi\rvert$ is large, block partitioning every odd number of sites will be needed to retain the antiferromagnetic order in the $x$ direction). Then, we diagonalize the intrablock Hamiltonian and project the Hilbert space onto the two-dimensional low-energy subspace in each block. Our real-space RG transformation is slightly different from that in Ref.~\cite{Langari2004}, in the sense that the intrablock Hamiltonian in our transformation includes the magnetic field only at the left site in each block while that in Ref.~\cite{Langari2004} includes the fields at both sites. An advantage of our scheme is that the critical point $\Gamma =\Gamma _\mathrm{c}$ and the critical exponent $\nu$ for $\Xi =0$ coincide with the exact results $\Gamma _\mathrm{c} =K$ and $\nu =1$, where $\nu$ is defined by the correlation length $L_\mathrm{corr} \sim\lvert\Gamma -\Gamma _\mathrm{c} \rvert ^{-\nu}$ near the critical point~\cite{Nishimori2011}.

We focus on the case where the magnitude of the $XX$ interactions is not large, or more precisely $\lvert\Xi\rvert <\sqrt{K^2 +\Gamma ^2}$. Since the overall energy scale is unimportant at zero temperature, we consider the ratios of the coupling constants $\gamma =\Gamma /K$ and $\xi =\Xi /K$. As derived in Appendix~\ref{sec.rgeqslargeu}, we find the RG equations
\begin{equation}
\gamma (l+1)=\gamma (l)^2 +\frac{\xi (l)(1+2\gamma (l)^2)}{\sqrt{1+\gamma (l)^2}} ,\quad\xi (l+1)=0 \label{eq.rsrg_rglargeu}
\end{equation}
for $l\geq 0$, where $\gamma (l)$ and $\xi (l)$ denote the coupling constants after $l$ RG steps.

It turns out that there are three fixed points $(\gamma ,\xi )=(0,0),(1,0),(\infty ,0)$. The two fixed points $(\gamma ,\xi )=(0,0),(\infty ,0)$ are stable and correspond to the staggered and symmetric phases in the original Ising ladder~\eqref{eq.model_hfrustisinglad}, respectively. The other fixed point $(\gamma ,\xi )=(1,0)$ is unstable under deviations in $\gamma$. Around this fixed point, $\gamma$ has the scaling dimension $y_\gamma =1$ since $\gamma (l+1)-1=\gamma (l)^2 -1\approx 2(\gamma (l)-1)$ for $l\geq 1$ and the present scaling factor is two. Thus, we have the critical exponent $\nu =1/y_\gamma =1$~\cite{Nishimori2011}.

Equation~\eqref{eq.rsrg_rglargeu} indicates that the $XX$ interaction is irrelevant in the limit $U\to\infty$ unless the bare magnitude $\lvert\Xi\rvert$ is large. This consequence is consistent with the result in Ref.~\cite{Langari2004}, which predicts that the antiferromagnetic $XX$ interaction gradually disappears when its bare magnitude is not large. It is also known for the model~\eqref{eq.model_hchain} with the antiferromagnetic $YY$ interactions of the magnitude $K$ that $\Xi$ is irrelevant for $\lvert\Xi /K\rvert\leq 1$ regardless of the sign of $\Xi$~\cite{Langari1998}.

\begin{figure}
\includegraphics{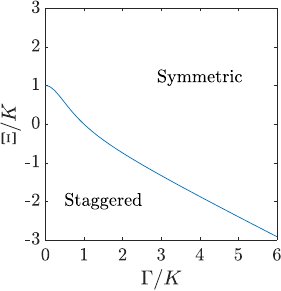}
\caption{Phase diagram of the Ising ladder in the limit $U\to\infty$ (the Ising chain) predicted by the real-space RG method. The blue line indicates second-order transition points.} \label{fig.rsrg_phasediagramlargeu}
\end{figure}

The critical points are determined by the equation
\begin{equation}
\gamma ^2 +\frac{\xi (1+2\gamma ^2)}{\sqrt{1+\gamma ^2}} =1\iff\xi =\frac{(1-\gamma ^2)\sqrt{1+\gamma ^2}}{1+2\gamma ^2} \label{eq.rsrg_cptlargeu}
\end{equation}
for the bare couplings $\gamma$ and $\xi$. Note that all the critical points but $(\gamma ,\xi )=(0,1)$ are inside the region $\lvert\xi\rvert <\sqrt{1+\gamma ^2}$, in which our analysis can be applied. Since the RG equations~\eqref{eq.rsrg_rglargeu} imply that the antiferromagnetic Ising chain with the transverse field and the $XX$ interactions belongs to the same universality class as the transverse-field Ising chain, there will be second-order transitions at the critical points~\eqref{eq.rsrg_cptlargeu}. We show the phase diagram in Fig.~\ref{fig.rsrg_phasediagramlargeu}, which indicates that the $\pm XX$-type catalysts do not eliminate the second-order phase transition encountered during quantum annealing. The phase diagram is in qualitative agreement with the diagrams obtained by Langari~\cite{Langari1998,Langari2004}, though the model in Ref.~\cite{Langari2004} covers only the case of the antiferromagnetic $XX$ interactions and that in Ref.~\cite{Langari1998} includes the $YY$ interactions that have the same sign and magnitude as the $ZZ$ interactions.

\section{DMRG calculation} \label{sec.dmrg}

We next employ the DMRG method~\cite{White1992,White1993,Schollwock2005,Hallberg2006} to obtain the phase diagram with a moderate coupling magnitude of $K/U$ in Sec.~\ref{ssec.dmrg_phasediagram}, which supplements the real-space RG analysis. In addition, we compute the energy gap of the finite-size system with a nonstoquastic $XX$ catalyst in Sec.~\ref{ssec.dmrg_energygap}.

\subsection{Phase diagram} \label{ssec.dmrg_phasediagram}

We perform the DMRG calculations for the frustrated Ising ladder with the transverse fields $\Gamma _\rmt =\Gamma _\rmb =\Gamma$ and the $XX$ interactions $\Xi _{\rmt\rmt} =\Xi$ and $\Xi _{\rmb\rmb} =\Xi _{\rmt\rmb} =0$ in Eq.~\eqref{eq.model_hfrustisinglad}.

To avoid trapping of the calculations in one of energy local minima, we start with several sweeps by taking not only the ground state but also the first excited state as the target states in the finite DMRG procedure, followed by the single target DMRG procedure to obtain the convergence of the ground-state energy. The truncation number $m$ of the density-matrix eigenvalues in the DMRG calculations is set as large as $m=200$ to achieve a maximal truncation error less than $10^{-12}$.

\begin{figure}
\includegraphics{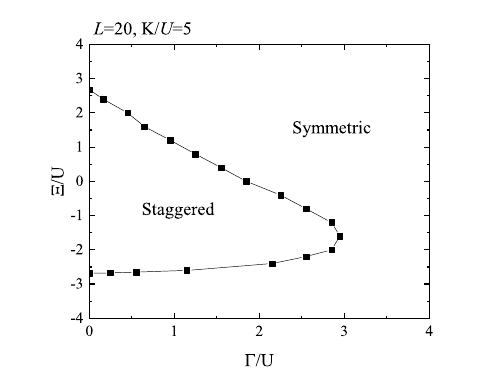}
\caption{Ground-state phase diagram of the frustrated Ising ladder obtained by the DMRG calculations with $K/U=5$ and $L=20$ under periodic boundary conditions. We set $\Gamma _\rmt =\Gamma _\rmb =\Gamma$, $\Xi _{\rmt\rmt} =\Xi$, and $\Xi _{\rmb\rmb} =\Xi _{\rmt\rmb} =0$ in Eq.~\eqref{eq.model_hfrustisinglad}. The staggered and symmetric phases are separated by a line of first-order transitions.} \label{fig.DMRG_phase1}
\end{figure}

We first show the ground-state phase diagram for $K/U=5$ and $L=20$ in Fig.~\ref{fig.DMRG_phase1}.\footnote{Note that the system size $L=20$ (40 spins) is far beyond the capacity of direct numerical diagonalization as adopted in Ref.~\cite{Laumann2012}, where the largest size was $L=12$.} Since the third condition of Eq.~\eqref{eq:non-stoq-cond} is satisfied, the Hamiltonian is stoquastic for $\Gamma =0\vee\Xi\geq 0$ and nonstoquastic for $\Gamma >0\wedge\Xi <0$. It is significant that the phase diagram obtained by the DMRG method has a similar shape to that for $K\to\infty$ in the thermodynamic limit $L\to\infty$ predicted by the real-space RG method (see Fig.~\ref{fig.rsrg_phasediagramlargek} for the latter diagram). As explained below, the phase boundary between the staggered phase and the symmetric phase in Fig.~\ref{fig.DMRG_phase1} is characterized by first-order transitions.

\begin{figure*}
\includegraphics{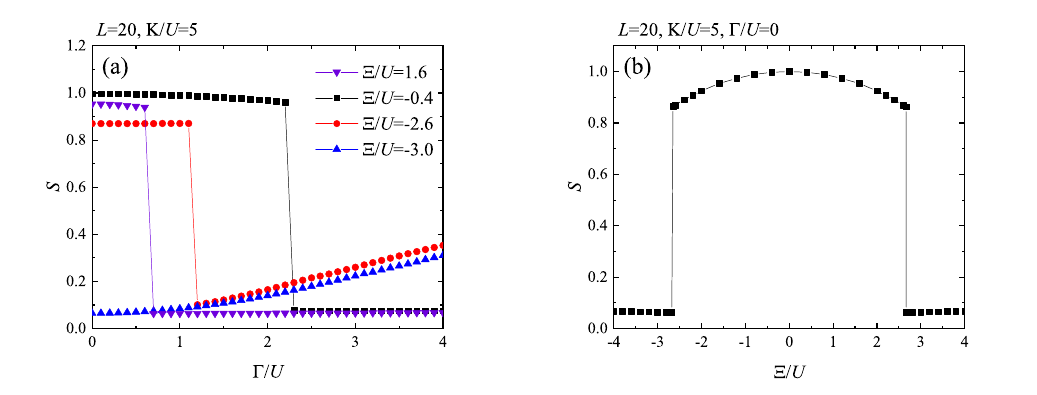}
\caption{Staggered order parameter on the top row $S$ (a) as a function of $\Gamma /U$ with several values of $\Xi /U$ and (b) as a function of $\Xi /U$ with $\Gamma=0$. The results are obtained by the DMRG method for $K/U=5$ and $L=20$ under periodic boundary conditions.} \label{fig.DMRG_OP}
\end{figure*}

In order to determine the phase boundary in Fig.~\ref{fig.DMRG_phase1}, we calculate the staggered order parameter for spins on the top row defined as
\begin{equation}
S=\frac{1}{L^2} \sum_{i=1}^L \sum_{j=1}^L (-1)^{i-j} \braket{\hat{Z}_{\rmt ,i} \hat{Z}_{\rmt ,j}}, \label{SOP}
\end{equation}
where $\braket{\cdots}$ denotes an average over the ground-state wave function. Figure~\ref{fig.DMRG_OP}(a) shows the results for $K/U=5$ and $L=20$ as a function of $\Gamma /U$ with several representative values of $\Xi /U$. When $\Xi /U=-0.4$, the staggered order parameter $S$ remains approximately $1$ for $\Gamma /U$ up to $\sim 2.2$, indicating that the ground state is in the staggered phase, and then suddenly decreases to almost zero ($S\approx 0.08$) as $\Gamma /U$ increases further, which corresponds to a first-order transition to the symmetric phase. A similar behavior continues until $\Xi /U\approx -2.65$ with varying $\Gamma /U$ as shown in Fig.~\ref{fig.DMRG_OP}(a), e.g., for $\Xi /U=-2.6$, where the first-order transition occurs at $\Gamma /U\approx 1.1$. On the other hand, when $\Xi /U=-3.0$, the staggered order parameter $S$ gradually increases from $S\approx 0.08$ with $\Gamma /U$, implying that the ground state remains in the symmetric phase. Indeed, as shown in Fig.~\ref{fig.DMRG_OP}(b), the first-order transition from the staggered phase to the symmetric phase occurs at $\Xi /U\approx\pm 2.65$ when $\Gamma =0$.

\begin{figure*}
\includegraphics{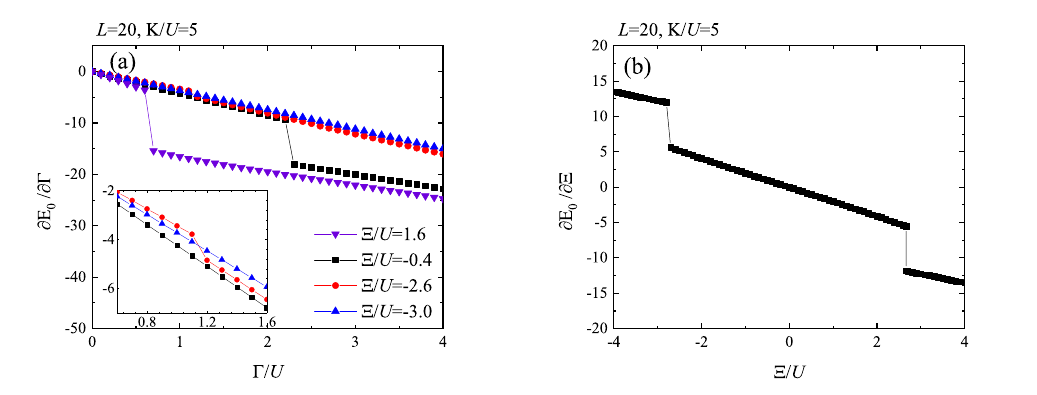}
\caption{First derivatives of the ground-state energy $E_0$ (a) with respect to $\Gamma$ for several values of $\Xi /U$ and (b) with respect to $\Xi$ for $\Gamma =0$. The results are obtained by the DMRG method for $K/U=5$ and $L=20$ under periodic boundary conditions. The inset in (a) is a magnification around $\Gamma /U=1$.} \label{fig.DMRG_FD}
\end{figure*}

To support these results, we calculate the first derivatives of the ground-state energy $E_0 =\braket{\hat{H}}$ with respect to $\Gamma$ and $\Xi$ since a first-order transition is characterized by a point where $\partial E_0 /\partial\Gamma$ or $\partial E_0 /\partial\Xi$ is discontinuous. According to the Hellmann-Feynman theorem~\cite{Feynman1939}, the first derivatives of the ground-state energy $E_0$ with respect to $\Gamma$ and $\Xi$ are calculated as
\bes
\begin{align}
\frac{\partial E_0}{\partial\Gamma} & =-\sum_{i=1}^L \braket{\hat X_{\rmt ,i} + \hat X_{\rmb ,i}} , \\
\frac{\partial E_0}{\partial\Xi} & =-\sum_{i=1}^L \braket{\hat X_{\rmt ,i} \hat X_{\rmt ,i+1}} .
\end{align}
\ees
As shown in Fig.~\ref{fig.DMRG_FD}, the first derivatives of the ground-state energy $E_0$ with respect to $\Gamma$ and $\Xi$ exhibit discontinuities exactly at the points where the staggered order parameter $S$ changes abruptly in Fig.~\ref{fig.DMRG_OP}, confirming the first-order nature of the transition.

\subsection{Energy gap} \label{ssec.dmrg_energygap}

\begin{figure}
\includegraphics{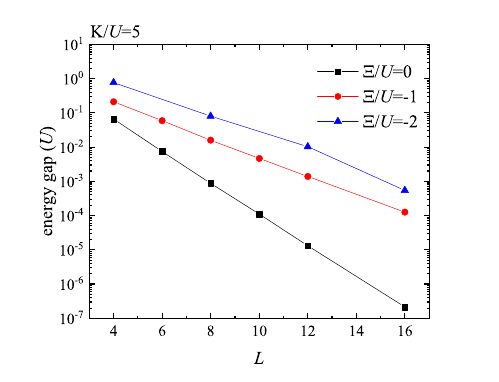}
\caption{The finite-size scaling of the minimum gap through the first-order transition driven by varying $\Gamma /U$ at a fixed value of $K/U=5$. The results are obtained by the DMRG method for three values of $\Xi /U$ under periodic boundary conditions.} \label{fig.DMRG_gap}
\end{figure}

We next obtain the energy gap between the ground state and the first excited state of the finite-size system with a nonstoquastic $XX$ catalyst using the DMRG method. We adopt the multi-target DMRG procedure with the ground state as well as the two lowest excited states as the target states, taking the truncation number $m$ as large as $2{,}000$. We show in Fig.~\ref{fig.DMRG_gap} the finite-size scaling of the minimum gap $\Delta (L)$ through the first-order transition driven by varying $\Gamma /U$ at $K/U=5$ for $\Xi /U=0,-1,-2$. We find that the gap $\Delta (L)$ decreases exponentially with $L$, i.e., $\ln\Delta (L)\sim -\alpha L$, for all three values of $\Xi /U$ (for $\Xi /U=0$ and $L\leq 12$, we reproduce the previously reported results in Ref.~\cite{Laumann2012}). It is noteworthy that the absolute value of the slope, $\alpha$, in $\ln\Delta (L)$ tends to be smaller in the presence of negative $\Xi$. From the point of view of quantum annealing and optimization this implies a quantitative, if not qualitative, improvement by the nonstoquastic $XX$ catalyst, since a larger gap implies a faster time-to-solution by the adiabatic theorem~\cite{Morita2008,Albash2018}.

We now discuss why the nonstoquastic $XX$ catalyst reduces the decay rate $\alpha$ of the minimum gap $\Delta (L)$. Consider the sum of the terms with the coefficients $U$, $\Gamma$, and $\Xi$ in the Hamiltonian $\hat{H}$ as a perturbation. It may be expected that the ground state and the first excited state at the transition are superpositions of columnar and staggered configurations and $\ln\Delta (L)$ is roughly proportional to $L\ln (\Gamma /K)$, because all the bottom spins need to be flipped by the bottom transverse field $\Gamma _\rmb =\Gamma$ to move from the symmetric phase to the staggered phase (note that the present $XX$ interactions $\Xi$ are applied only along the top row and cannot flip the bottom spins). Indeed, $\log _{10} (1.85/5)=-0.43$ and $\log _{10} (2.85/5)=-0.24$ are close to the slopes of the lines for $\Xi /U=0,-2$ in Fig.~\ref{fig.DMRG_gap}, respectively, where $\Gamma /U=1.85$ is the transition point for $\Xi /U=0$ and $\Gamma /U=2.85$ is for $\Xi /U=-2$ when $K/U=5$ (see Fig.~\ref{fig.DMRG_phase1}). This is consistent with the fact that the decay rate $\alpha$ is proportional to $\ln (K/\Gamma )$ when $K/U$ is varied in the absence of the $XX$ catalyst~\cite{Laumann2012}. Our argument suggests that the nonstoquastic $XX$ catalyst on the top row with an appropriate magnitude $\Xi$ increases the transverse field $\Gamma$ at the transition and reduces the exponential decay rate of the gap in the present system. Although we have not carried out a numerical calculation of the gap of the model with the stoquastic $XX$ catalyst due to the associated heavy computational cost, it is implied that the decay rate $\alpha$ of the minimum gap $\Delta (L)$ becomes larger for $\Xi >0$ (as well as for $\Xi /U\lesssim -2.5$) than for $\Xi =0$, because $\Gamma /U$ at the transition is smaller in the former case, as shown in Fig.~\ref{fig.DMRG_phase1}.

Similar arguments were presented for a geometrically local Ising model on two connected rings~\cite{Albash2019}. There it was shown by numerical diagonalization that the nonstoquastic $XX$ catalyst makes the minimum gap larger than in the case without the $XX$ catalyst and in the case with the stoquastic $XX$ catalyst. It was argued that one of the possible reasons for the softening of the avoided level crossing with the nonstoquastic catalyst is that the driver with the nonstoquastic $XX$ catalyst causes the level crossing earlier in quantum annealing (which means a larger transverse field) than with the stoquastic $XX$ catalyst, which is a consequence of perturbation theory.

\section{Summary and Conclusions} 
\label{sec.conc}

We have studied the effects of stoquastic and nonstoquastic catalysts, implemented via ferromagnetic and antiferromagnetic $XX$ interactions, in the setting of the frustrated Ising ladder. This model -- without the $XX$ interactions -- is known to have a first-order phase transition with an exponentially decaying 
energy gap, which is characterized by a change in the topology of dimer configurations in the limit of strong frustration $K\to\infty$~\cite{Laumann2012}. We have formulated a real-space RG transformation such that the symmetry of the problem is preserved and 
used it to obtain the phase diagram in the presence of $XX$ interactions of both signs, stoquastic and nonstoquastic. The result shows that the first-order transition persists in the presence of $XX$ interactions of moderate magnitude. The transition point obtained by the real-space RG method in the case without $XX$ interactions is close to the value obtained by numerical diagonalization~\cite{Laumann2012}. This is surprising, given that the real-space RG approach involves a number of uncontrolled approximations. In addition, we applied the real-space RG method to the case with small frustration and found that the second-order transition persists under the influence of $XX$ interactions of moderate magnitude.

We next performed extensive numerical computations by the DMRG method for a large but finite value of $K$ in order to directly study the behavior of various physical quantities. The results for the order parameter and derivatives of the ground-state energy clearly indicate the existence of first-order phase transitions in the presence of stoquastic or nonstoquastic $XX$ interactions, which is consistent with the conclusion from the real-space RG study. The structure of the phase diagram qualitatively and even semi-quantitatively resembles the one obtained by the real-space RG analysis. Our DMRG results furthermore confirm that the energy gap decays exponentially as a function of system size at the first-order transition points, both with or without  $XX$ interactions on the top row of the ladder. However, the nonstoquastic $XX$ catalyst reduces the decay rate of the gap, as we showed numerically. We pointed out that this softening of the exponential decay of the gap may be due to an increase of the transverse field at the transition in the presence of a nonstoquastic catalyst.

A first-order phase transition is a sudden change of the system state between very different phases, e.g., between water and ice, and is unlikely to be induced or reduced by a series of gradual local changes. In the case of spin systems, the latter gradual change is exemplified by the introduction of $XX$ interactions, which change the state of the system in the computational basis by flipping only pairs of spins simultaneously. Nevertheless, there exist examples in which nonstoquastic $XX$ interactions change the order of a phase transition from first to second~\cite{Seki2012,Seoane2012,Seki2015,Nishimori2017,Albash2019,Takada2020}, meaning a drastic reduction of the ``strength'' of the phase transition by nonstoquastic $XX$ interactions, although counterexamples abound~\cite{Crosson2020}.

The first-order transition in the problem of the frustrated Ising ladder without $XX$ interactions belongs to a more stable class in the sense that the two phases are separated by different topological structures in the limit of strong frustration. These topological structures are generated by the columnar and staggered configurations, which have magnetizations with opposite signs on the bottom row of the ladder. In the presence of frustration of infinite magnitude, the addition of any local operators of finite magnitude does not allow transitions between the topologically distinct states. It is therefore expected that the first-order transition is stable against the introduction of the $XX$ interactions of either sign.

From the perspective of quantum annealing, the persistence of the first-order transition means that the computational complexity of the problem, exponential in the system size, remains intact under the introduction of stoquastic or nonstoquastic $XX$ catalysts as long as the system evolves under adiabatic unitary dynamics. It is an important and interesting open question to study whether any of these conclusions are modified under nonunitary (open-system) dynamics or diabatic evolution, both of which take place in real quantum devices~\cite{crosson2020prospects}.

\begin{acknowledgments}
The work of KT was supported by JSPS KAKENHI Grant No.~17J09218.
The research is based upon work partially supported by the Office of the Director of National Intelligence (ODNI), Intelligence Advanced Research Projects Activity (IARPA) and the Defense Advanced Research Projects Agency (DARPA), via the U.S. Army Research Office contract W911NF-17-C-0050. The views and conclusions contained herein are those of the authors and should not be interpreted as necessarily representing the official policies or endorsements, either expressed or implied, of the ODNI, IARPA, DARPA, or the U.S. Government. The U.S. Government is authorized to reproduce and distribute reprints for Governmental purposes notwithstanding any copyright annotation thereon.
\end{acknowledgments}

\appendix

\section{Energy gap of the model without $XX$ terms} \label{app:gap}

\begin{figure}
\includegraphics{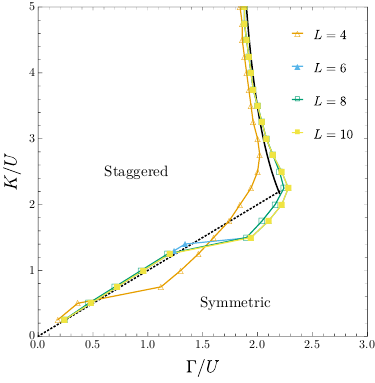}
\caption{Locations of the minimum energy gap of the Ising ladder with uniform transverse field $\Gamma$ and no $XX$ terms for several system sizes $L$, at fixed values of $K/U$. The first- and second-order phase boundaries predicted by perturbation theory are indicated by black solid and dashed lines, respectively.} \label{fig.minimum_gap}
\end{figure}

This appendix displays additional data to supplement the phase diagram of the Ising ladder without $XX$ terms in Fig.~\ref{fig.phase_diagram_original}. As mentioned in Sec.~\ref{ssec.model_phasediagramwoxx}, it appears that the minimum gap point (namely, the location at which the energy gap between the ground state and the first excited state is minimized for a fixed $K/U$) has a discontinuity at $K/U\approx 1.5$ between $\Gamma /U\approx 1.5$ and $1.9$. To examine whether this phenomenon is a finite-size effect, we plot the minimum gap points for $L=4,6,8,10$ in Fig.~\ref{fig.minimum_gap}. We find near convergence already for $L=10$, which suggests that the discontinuity of the minimum gap point location is not a finite size effect.

\begin{figure}
\includegraphics{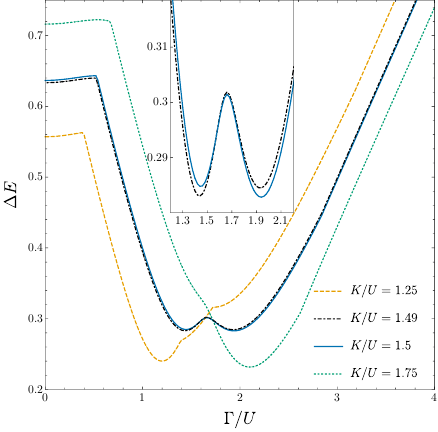}
\caption{Energy gap $\Delta E$ as a function of $\Gamma /U$ for $L=10$ at different values of $K/U$. The inset is a magnification of the lines for $K/U=1.49,1.5$, which shows that the global minimum of $\Delta E$ moves from the left ``well'' to the right ``well'' as $K/U$ changes.} \label{fig.gap_vs_Gamma}
\end{figure}

An explanation is provided in Fig.~\ref{fig.gap_vs_Gamma}, which shows the energy gap $\Delta E$ as a function of $\Gamma /U$ for $L=10$ and $K/U=1.25,1.49,1.5,1.75$. It turns out that the energy gap for $K/U\approx 1.5$ takes a ``double-well'' form, which is the origin of the discontinuity of the minimum gap point. The energy gap $\Delta E$ is minimized in the left ``well'' for $K/U\lesssim 1.5$, while $\Delta E$ is minimized in the right ``well'' for $K/U\gtrsim 1.5$. We leave a detailed understanding of this phenomenon as a future topic of research.

\begin{widetext}

\section{Derivation of RG equations in the limit of large frustration} \label{sec.rgeqslargek}

We derive the RG equations of the frustrated Ising ladder in the limit of large frustration (i.e., the dimer limit), which were used in Sec.~\ref{ssec.rsrg_largek}. Appendix~\ref{ssec.rgeqslargek_preanl} gives a review of basic properties of the model. In Appendix~\ref{ssec.rgeqslargek_hamiltonian}, we define a Hamiltonian that appears in the RG analysis. In Appendices~\ref{ssec.rgeqslargek_blockpart}--\ref{ssec.rgeqslargek_rgeq}, we explain the way to construct the real-space RG transformation including the variational ansatz in detail and write down the RG equations. We also calculate the renormalized couplings and show their behavior.

\subsection{Preliminary analysis} \label{ssec.rgeqslargek_preanl}

To prepare for the real-space RG analysis, we demonstrate why the columnar and staggered configurations are the low-energy states of the frustrated Ising ladder~\eqref{eq.model_hfrustisinglad} in the dimer limit $K\to\infty$ and derive the number of low-energy states, as indicated in Ref.~\cite{Laumann2012}.

Since the overall energy scale does not change the statistical properties in the zero-temperature limit, we consider the dimensionless Hamiltonian $\hat{h} =\hat{H}/K$:
\begin{align}
\hat{h} & =\sum _{i=1}^L \Bigl(\hat{Z}_{\rmt ,i} \hat{Z}_{\rmt ,i+1} -\hat{Z}_{\rmb ,i} \hat{Z}_{\rmb ,i+1} -\hat{Z}_{\rmt ,i} \hat{Z}_{\rmb ,i} -\hat{Z}_{\rmt ,i} \notag \\
& \hphantom{{} =\sum\Bigl(} +\frac{u}{2} \hat{Z}_{\rmb ,i} -(\gamma _\rmt \hat{X}_{\rmt ,i} +\gamma _\rmb \hat{X}_{\rmb ,i})-(\xi _{\rmt\rmt} \hat{X}_{\rmt ,i} \hat{X}_{\rmt ,i+1} +\xi _{\rmb\rmb} \hat{X}_{\rmb ,i} \hat{X}_{\rmb ,i+1} +\xi _{\rmt\rmb} \hat{X}_{\rmt ,i} \hat{X}_{\rmb ,i})\Bigr) ,
\end{align}
where $u=U/K$, $\gamma _a =\Gamma _a /K$, and $\xi _{aa'} =\Xi _{aa'} /K$ are sufficiently small dimensionless parameters. We assume that $u$, $\gamma _a$, and $\xi _{aa'}$ are of the same order $\delta\ll 1$. We separate the Hamiltonian $\hat{h}$ into $\hat{h}^{(0)} =\mathcal{O} (\delta ^0)$ and $\hat{h}^{(1)} =\mathcal{O} (\delta ^1)$:
\bes
\label{eq.rgeqslargek_origh-all}
\begin{align}
\hat{h} & =\hat{h}^{(0)} +\hat{h}^{(1)} , \label{eq.rgeqslargek_origh} \\
\hat{h}^{(0)} & =\sum _{i=1}^L (\hat{Z}_{\rmt ,i} \hat{Z}_{\rmt ,i+1} -\hat{Z}_{\rmb ,i} \hat{Z}_{\rmb ,i+1} -\hat{Z}_{\rmt ,i} \hat{Z}_{\rmb ,i} -\hat{Z}_{\rmt ,i}), \label{eq.rgeqslargek_origh0} \\
\hat{h}^{(1)} & =\sum _{i=1}^L \left(\frac{u}{2} \hat{Z}_{\rmb ,i} -(\gamma _\rmt \hat{X}_{\rmt ,i} +\gamma _\rmb \hat{X}_{\rmb ,i})-(\xi _{\rmt\rmt} \hat{X}_{\rmt ,i} \hat{X}_{\rmt ,i+1} +\xi _{\rmb\rmb} \hat{X}_{\rmb ,i} \hat{X}_{\rmb ,i+1} +\xi _{\rmt\rmb} \hat{X}_{\rmt ,i} \hat{X}_{\rmb ,i})\right) . \label{eq.rgeqslargek_origh1}
\end{align}
\ees

Using translational invariance yields $\sum _i \hat{Z}_{\rmt ,i} =\sum _i (\hat{Z}_{\rmt ,i} +\hat{Z}_{\rmt ,i+1}) / 2$ and $\sum _i \hat{Z}_{\rmt ,i} \hat{Z}_{\rmb ,i} = \sum _i (\hat{Z}_{\rmt ,i} \hat{Z}_{\rmb ,i} +\hat{Z}_{\rmt ,i+1} \hat{Z}_{\rmb ,i+1}) / 2$. We can thus rewrite the zeroth-order Hamiltonian $\hat{h}^{(0)}$ as
\begin{align}
\hat{h}^{(0)} =\sum _{i=1}^L \left(\hat{Z}_{\rmt ,i} \hat{Z}_{\rmt ,i+1} -\hat{Z}_{\rmb ,i} \hat{Z}_{\rmb ,i+1} -\hat{Z}_{\rmt ,i} \frac{1+\hat{Z}_{\rmb ,i}}{2} -\hat{Z}_{\rmt ,i+1} \frac{1+\hat{Z}_{\rmb ,i+1}}{2} \right) .
\end{align}
Next, let us rewrite the same Hamiltonian as a linear combination of projectors:
\begin{align}
\hat{h}^{(0)} & =\sum _{i=1}^L \Biggl({} %
-2\Ket{\stack{\uar\uar\\ \uar\uar}} \Bra{\stack{\uar\uar\\ \uar\uar}} -2\Ket{\stack{\dar\uar\\ \uar\uar}} \Bra{\stack{\dar\uar\\ \uar\uar}} -2\Ket{\stack{\uar\dar\\ \uar\uar}} \Bra{\stack{\uar\dar\\ \uar\uar}} +2\Ket{\stack{\dar\dar\\ \uar\uar}} \Bra{\stack{\dar\dar\\ \uar\uar}} \notag \\
& \hphantom{{} =\sum\Biggl(} %
+0\Ket{\stack{\uar\uar\\ \dar\dar}} \Bra{\stack{\uar\uar\\ \dar\dar}} -2\Ket{\stack{\dar\uar\\ \dar\dar}} \Bra{\stack{\dar\uar\\ \dar\dar}} -2\Ket{\stack{\uar\dar\\ \dar\dar}} \Bra{\stack{\uar\dar\\ \dar\dar}} +0\Ket{\stack{\dar\dar\\ \dar\dar}} \Bra{\stack{\dar\dar\\ \dar\dar}} \notag \\
& \hphantom{{} =\sum\Biggl(} %
+1\Ket{\stack{\uar\uar\\ \dar\uar}} \Bra{\stack{\uar\uar\\ \dar\uar}}
-1\Ket{\stack{\dar\uar\\ \dar\uar}} \Bra{\stack{\dar\uar\\ \dar\uar}}
+1\Ket{\stack{\uar\dar\\ \dar\uar}} \Bra{\stack{\uar\dar\\ \dar\uar}}
+3\Ket{\stack{\dar\dar\\ \dar\uar}} \Bra{\stack{\dar\dar\\ \dar\uar}} \notag \\
& \hphantom{{} =\sum\Biggl(} %
+1\Ket{\stack{\uar\uar\\ \uar\dar}} \Bra{\stack{\uar\uar\\ \uar\dar}}
+1\Ket{\stack{\dar\uar\\ \uar\dar}} \Bra{\stack{\dar\uar\\ \uar\dar}}
-1\Ket{\stack{\uar\dar\\ \uar\dar}} \Bra{\stack{\uar\dar\\ \uar\dar}}
+3\Ket{\stack{\dar\dar\\ \uar\dar}} \Bra{\stack{\dar\dar\\ \uar\dar}} \Biggr) _{i,i+1} .
\end{align}
Here, each state vector containing an array of arrows means a product state and the positions in the array correspond to those in the ladder:
\begin{equation}
\Ket{\stack{\sigma _{\rmt ,i} \\ \sigma _{\rmb ,i}} \stack{\sigma _{\rmt ,i+1} \\ \sigma _{\rmb ,i+1}}}_{i,i+1} :=\ket{\sigma _{\rmt ,i}}_{\rmt ,i} \ket{\sigma _{\rmb ,i}}_{\rmb ,i} \ket{\sigma _{\rmt ,i+1}}_{\rmt ,i+1} \ket{\sigma _{\rmb ,i+1}}_{\rmb ,i+1} \quad (\sigma _{a,j} =\uar ,\dar ).
\end{equation}
We denoted the normalized eigenstates of $\hat{Z}_{a,i}$ with the eigenvalues $+1$ and $-1$ by $\ket{\uar}_{a,i}$ and $\ket{\dar}_{a,i}$, respectively.

Since a constant energy difference is unimportant, we add $2\hat{I}$ to $\hat{h}^{(0)}$, where $\hat{I}$ is the identity operator, and redefine $\hat{h}^{(0)} +2\hat{I}$ as $\hat{h}^{(0)}$:
\begin{align}
\hat{h}^{(0)} & =\sum _{i=1}^L \Biggl({} %
+4\Ket{\stack{\dar\dar\\ \uar\uar}} \Bra{\stack{\dar\dar\\ \uar\uar}} +2\Ket{\stack{\uar\uar\\ \dar\dar}} \Bra{\stack{\uar\uar\\ \dar\dar}} +2\Ket{\stack{\dar\dar\\ \dar\dar}} \Bra{\stack{\dar\dar\\ \dar\dar}} \notag \\
& \hphantom{{} =\sum\Biggl(} %
+3\Ket{\stack{\uar\uar\\ \dar\uar}} \Bra{\stack{\uar\uar\\ \dar\uar}}
+1\Ket{\stack{\dar\uar\\ \dar\uar}} \Bra{\stack{\dar\uar\\ \dar\uar}}
+3\Ket{\stack{\uar\dar\\ \dar\uar}} \Bra{\stack{\uar\dar\\ \dar\uar}}
+5\Ket{\stack{\dar\dar\\ \dar\uar}} \Bra{\stack{\dar\dar\\ \dar\uar}} \notag \\
& \hphantom{{} =\sum\Biggl(} %
+3\Ket{\stack{\uar\uar\\ \uar\dar}} \Bra{\stack{\uar\uar\\ \uar\dar}}
+3\Ket{\stack{\dar\uar\\ \uar\dar}} \Bra{\stack{\dar\uar\\ \uar\dar}}
+1\Ket{\stack{\uar\dar\\ \uar\dar}} \Bra{\stack{\uar\dar\\ \uar\dar}}
+5\Ket{\stack{\dar\dar\\ \uar\dar}} \Bra{\stack{\dar\dar\\ \uar\dar}} \Biggr) _{i,i+1} . \label{eq.rgeqslargek_origh0shifted}
\end{align}
These terms can be regarded as energy penalties for the 11 configurations of each pair $(i,i+1)$. A state that does not pay a penalty for any pair $(i,i+1)$ is a ground state of $\hat{h}^{(0)}$ if such a state exists. We find that the ground states of $\hat{h}^{(0)}$ are the columnar and staggered configurations defined in Sec.~\ref{sec.model} and their superpositions, on which no energy penalties are imposed. In other words, the ground space of $\hat{h}^{(0)}$ is given by $\mathcal{D}_L =\Span\{\,\ket{\sigma} \,\} _{\sigma\in D_L}$, where
\begin{align}
D_L & :=\left\{\left(\stack{\sigma _{\rmt ,1} \\ \uar} \stack{\dotsm\\ \dotsm} \stack{\sigma _{\rmt ,L} \\ \uar} \right)\bigg\vert\,\sigma _{\rmt ,1} ,\dots ,\sigma _{\rmt ,L} \in\{\uar ,\dar\}\wedge (\sigma _{\rmt ,1} ,\sigma _{\rmt ,2}),(\sigma _{\rmt ,2} ,\sigma _{\rmt ,3}),\dots ,(\sigma _{\rmt ,L-1} ,\sigma _{\rmt ,L}),(\sigma _{\rmt ,L} ,\sigma _{\rmt ,1})\not= (\dar ,\dar)\right\} \notag \\
& \hphantom{{} := {}} \cup\left\{\left(\stack{\dar\uar\dotsm\dar\uar\\ \dar\dar\dotsm\dar\dar} \right) ,\left(\stack{\uar\dar\dotsm\uar\dar\\ \dar\dar\dotsm\dar\dar} \right)\right\} .
\end{align}
Although the full Hilbert space is $\mathcal{H}_L =\Span\{\,\ket{\sigma} \,\} _{\sigma\in H_L}$ with
\begin{equation}
H_L :=\left\{\left(\stack{\sigma _{\rmt ,1} \\ \sigma _{\rmb ,1}} \stack{\dotsm\\ \dotsm} \stack{\sigma _{\rmt ,L} \\ \sigma _{\rmb ,1}} \right)\bigg\vert\,\text{$\sigma _{\rmt ,1} ,\dots ,\sigma _{\rmt ,L} ,\sigma _{\rmb ,1} ,\dots ,\sigma _{\rmb ,L} \in\{\uar ,\dar\}$} \right\} ,
\end{equation}
$\hat{h}^{(0)}$ imposes a penalty on every state in $\mathcal{H}_L \setminus\mathcal{D}_L$. As mentioned in Sec.~\ref{sec.model}, the elements of $D_L$ can be assigned to dimer coverings on a two-leg ladder.

What is the dimension of the nonpenalized subspace, $\dim\mathcal{D}_L =\lvert D_L \rvert$? First consider the open-ladder counterpart of $D_L$:
\begin{align}
D'_L & :=\left\{\left(\stack{\sigma _{\rmt ,1} \\ \uar} \stack{\dotsm\\ \dotsm} \stack{\sigma _{\rmt ,L} \\ \uar} \right)\bigg\vert\,\sigma _{\rmt ,1} ,\dots ,\sigma _{\rmt ,L} \in\{\uar ,\dar\}\wedge (\sigma _{\rmt ,1} ,\sigma _{\rmt ,2}),(\sigma _{\rmt ,2} ,\sigma _{\rmt ,3}),\dots ,(\sigma _{\rmt ,L-1} ,\sigma _{\rmt ,L})\not= (\dar ,\dar)\right\} \notag \\
& \hphantom{{} := {}} \cup\left\{\left(\stack{\dar\uar\dotsm\\ \dar\dar\dotsm} \right) ,\left(\stack{\uar\dar\dotsm\\ \dar\dar\dotsm} \right)\right\} .
\end{align}
In particular, we focus on the columnar configurations:
\begin{equation}
D'^{\,\text{columnar}}_L :=\left\{\left(\stack{\sigma _{\rmt ,1} \\ \uar} \stack{\dotsm\\ \dotsm} \stack{\sigma _{\rmt ,L} \\ \uar} \right)\bigg\vert\,\sigma _{\rmt ,1} ,\dots ,\sigma _{\rmt ,L} \in\{\uar ,\dar\}\wedge (\sigma _{\rmt ,1} ,\sigma _{\rmt ,2}),(\sigma _{\rmt ,2} ,\sigma _{\rmt ,3}),\dots ,(\sigma _{\rmt ,L-1} ,\sigma _{\rmt ,L})\not= (\dar ,\dar)\right\} .
\end{equation}
The number of elements of this set $F_L :=\lvert D'^{\,\text{columnar}}_L \rvert$ satisfies
\begin{equation}
F_1 =2,\quad F_2 =3,\quad F_L =F_{L-1} +F_{L-2} , \label{eq.rgeqslargek_fibonacci}
\end{equation}
which means that $F_L$ are the Fibonacci numbers (note the values of $F_1$ and $F_2$). We can obtain the recurrence relation for the following reason: If $\sigma _{\rmt ,L}$ is up, $(\sigma _{\rmt ,1} ,\dots ,\sigma _{\rmt ,L-1})$ can be regarded as an open chain of length $L-1$. On the other hand, if $\sigma _{\rmt ,L}$ is down, $\sigma _{\rmt ,L-1}$ should be up and $(\sigma _{\rmt ,1} ,\dots ,\sigma _{\rmt ,L-2})$ is an open chain of length $L-2$.

We can express the dimension of the nonpenalized subspace for the original periodic ladder $\lvert D_L\rvert$ using the Fibonacci numbers. If the bottom spins are up, setting $\sigma _{\rmt ,L} =\uar$ yields an open chain of length $L-1$ while setting $\sigma _{\rmt ,L} =\dar$ fixes $\sigma _{\rmt ,L-1}$ and $\sigma _{\rmt ,1}$ to up and yields an open chain of length $L-3$. If the bottom spins are down, the top spins can take the two antiferromagnetic configurations. Thus, we have
\begin{equation}
\lvert D_L \rvert =F_{L-1} +F_{L-3} +2.
\end{equation}
Since the Fibonacci sequence asymptotically behaves as
\begin{equation}
F_L \sim\varphi ^L \quad (L\to\infty )
\end{equation}
with $\varphi =(1+\sqrt{5})/2$ being the golden ratio, $\lvert D_L \rvert$ is exponentially large as a function of $L$. Thus, the ground state of $\hat{h}^{(0)}$ is exponentially degenerate.

\subsection{Generalized Hamiltonian} \label{ssec.rgeqslargek_hamiltonian}

The bare dimensionless Hamiltonian is given by Eqs.~\eqref{eq.rgeqslargek_origh}, \eqref{eq.rgeqslargek_origh0shifted}, and \eqref{eq.rgeqslargek_origh1}. Now we define a more general Hamiltonian, because an RG transformation will produce interactions not included in the bare Hamiltonian:
\bes
\label{eq.rgeqslargek_hall}
\begin{align}
\hat{h} & =\hat{h}^{(0)} +\hat{h}^{(1)} , \label{eq.rgeqslargek_h} \\
\hat{h}^{(0)} & =\sum _{i=1}^L \sum _{(\sigma _i ,\sigma _{i+1})\in H_2 \setminus D'_2} k(\sigma _i ,\sigma _{i+1})(\ket{\sigma _i \sigma _{i+1}} \bra{\sigma _i \sigma _{i+1}})_{i,i+1} , \label{eq.rgeqslargek_h0} \\
\hat{h}^{(1)} & =\sum _{i=1}^L \left[\frac{u}{2} \hat{Z}_{\rmb ,i} -\gamma\hat{X}_{\rmt ,i} \frac{1+\hat{Z}_{\rmb ,i}}{2} +v\hat{Z}_{\rmt ,i} \frac{1+\hat{Z}_{\rmb ,i}}{2} -\xi\left(\Ket{\stack{\uar\dar\\ \uar\uar}} \Bra{\stack{\dar\uar\\ \uar\uar}} +\Ket{\stack{\dar\uar\\ \uar\uar}} \Bra{\stack{\uar\dar\\ \uar\uar}} \right) _{i,i+1} \right] +\hat{o} \hat{P} (\mathcal{D}_L^\perp )+\hat{P} (\mathcal{D}_L^\perp )\hat{o}^\dagger , \label{eq.rgeqslargek_h1}
\end{align}
\ees
where $0<k(\sigma _i ,\sigma _{i+1})=\mathcal{O} (\delta ^0)$ for all $(\sigma _i ,\sigma _{i+1})\in H_2 \setminus D'_2$ and $u,\gamma ,v,\xi =\mathcal{O} (\delta ^1)$. The length $L$ is even and position $L+1$ is identified with $1$ due to the assumption of periodic boundary conditions. We denoted an arbitrary operator of $\mathcal{O} (\delta ^1)$ by $\hat{o}$ and the projection operator onto $\mathcal{D}_L^\perp$ by $\hat{P} (\mathcal{D}_L^\perp )$, where $\mathcal{D}_L^\perp$ is the orthogonal complement of $\mathcal{D}_L$. The operator $\hat{o} \hat{P} (\mathcal{D}_L^\perp )+\hat{P} (\mathcal{D}_L^\perp )\hat{o}^\dagger$ in $\hat{h}^{(1)}$ will disappear after an RG transformation and is irrelevant in the sense of RG theory. The bare Hamiltonian can be obtained by assigning the coefficients in Eq.~\eqref{eq.rgeqslargek_origh0shifted} to $k(\sigma _i ,\sigma _{i+1})$ and setting $\gamma =\gamma _\rmt$, $v=0$, and $\xi =\xi _{\rmt\rmt}$.

In Eq.~\eqref{eq.rgeqslargek_h0}, we denoted configurations of two spins at each position $i$ by
\begin{equation}
\sigma _i \in H_1 =\left\{\left(\stack{\uar\\ \uar} \right) ,\left(\stack{\dar\\ \uar} \right) ,\left(\stack{\uar\\ \dar} \right) ,\left(\stack{\dar\\ \dar} \right)\right\} .
\end{equation}
Every configuration $(\sigma _i ,\sigma _{i+1})\in H_2 \setminus D'_2$ costs energy of $\mathcal{O} (\delta ^0)$ for each pair $(i,i+1)$, where $D'_2$ and $H_2 \setminus D'_2$ are given by
\bes
\begin{align}
D'_2 & =\left\{\left(\stack{\uar\uar\\ \uar\uar} \right) ,\left(\stack{\dar\uar\\ \uar\uar} \right) ,\left(\stack{\uar\dar\\ \uar\uar} \right) ,\left(\stack{\dar\uar\\ \dar\dar} \right) ,\left(\stack{\uar\dar\\ \dar\dar} \right)\right\} , \\
H_2 \setminus D'_2 & =\Biggl\{\left(\stack{\dar\dar\\ \uar\uar} \right) ,\left(\stack{\uar\uar\\ \dar\dar} \right) ,\left(\stack{\dar\dar\\ \dar\dar} \right) ,\left(\stack{\uar\uar\\ \dar\uar} \right) ,\left(\stack{\dar\uar\\ \dar\uar} \right) ,\left(\stack{\uar\dar\\ \dar\uar} \right) ,\left(\stack{\dar\dar\\ \dar\uar} \right) ,\left(\stack{\uar\uar\\ \uar\dar} \right) ,\left(\stack{\dar\uar\\ \uar\dar} \right) ,\left(\stack{\uar\dar\\ \uar\dar} \right) ,\left(\stack{\dar\dar\\ \uar\dar} \right)\Biggr\} .
\end{align}
\ees
Since $(\sigma _i ,\sigma _{i+1})\in H_2 \setminus D'_2$ ($\forall i=1,\dots ,L$) is equivalent to $(\sigma _1 ,\dots ,\sigma _L)\in H_L \setminus D_L$, any $\ket{\psi} \in\mathcal{H}_L \setminus\mathcal{D}_L$ has a nonvanishing zeroth-order energy expectation value $0<\braket{\psi |\hat{h}^{(0)}|\psi} =\mathcal{O} (\delta ^0)$. Hence, the zeroth-order Hamiltonian $\hat{h}^{(0)}$ imposes an energy penalty on every state in $\mathcal{H}_L \setminus\mathcal{D}_L$.

\subsection{Block partition} \label{ssec.rgeqslargek_blockpart}

As a first step of an RG transformation, we partition the ladder into $\widetilde{L} =L/b$ blocks of spins, where $b$ is the scaling factor. We project the Hilbert space onto a four-dimensional space for each block. The projector is written as
\bes
\begin{align}
\hat{Q} & =\bigotimes _{I=1}^{\widetilde{L}} \hat{Q}_I =\sum _{\sigma\in H_{\widetilde{L}}} \widetilde{\ket{\sigma}} \widetilde{\bra{\sigma}} , \\
\hat{Q}_I & =\left(\widetilde{\Ket{\stack{\uar\\ \uar}}} \widetilde{\Bra{\stack{\uar\\ \uar}}} +\widetilde{\Ket{\stack{\dar\\ \uar}}} \widetilde{\Bra{\stack{\dar\\ \uar}}} +\widetilde{\Ket{\stack{\uar\\ \dar}}} \widetilde{\Bra{\stack{\uar\\ \dar}}} +\widetilde{\Ket{\stack{\dar\\ \dar}}} \widetilde{\Bra{\stack{\dar\\ \dar}}} \right) _I ,
\end{align}
\ees
where $\widetilde{\ket{\sigma}} =\bigotimes _{I=1}^{\widetilde{L}} \widetilde{\ket{\sigma _I}}_I$ for $\sigma =(\sigma _1 ,\dots ,\sigma _{\widetilde{L}})$ and $\{\,\widetilde{\ket{\sigma _I}}_I \,\} _{\sigma _I \in H_1}$ is a set of four orthonormal states in the $I$th block constituted by the $2b$ sites $(a,i)$ ($a=\rmt ,\rmb$; $i=b(I-1)+1,\dots ,bI$). In the present system, the scaling factor $b$ should be an odd number to prevent the RG transformation from breaking the antiferromagnetic order in the staggered phase (an RG transformation with even $b$ would turn antiferromagnetic order into ferromagnetic order). We take $b=3$ in the following.

Each $\widetilde{\ket{\sigma _I}}_I$ is a superposition of $\ket{\tau _{3I-2} \tau _{3I-1} \tau _{3I}}_I =\ket{\tau _{3I-2}}_{3I-2} \ket{\tau _{3I-1}}_{3I-1} \ket{\tau _{3I}}_{3I}$ ($(\tau _{3I-2} ,\tau _{3I-1} ,\tau _{3I})\in H_3$). We suppose that for each $\sigma _1 \in H_1$, $\widetilde{\ket{\sigma _1}}_I$ take the same form except for the difference of the block. In other words, when we express $\widetilde{\ket{\sigma _1}}_I$ as a linear combination of $\ket{\tau _1 \tau _2 \tau _3}_I$, the coefficients are independent of $I$.

We expand the states $\widetilde{\ket{\sigma _I}}_I$ ($\sigma _I \in H_1$) up to first order in powers of $\delta$:
\begin{equation}
\widetilde{\ket{\sigma _I}}_I =\widetilde{\ket{\sigma _I}}_I^{(0)} +\widetilde{\ket{\sigma _I}}_I^{(1)} +\mathcal{O} (\delta ^2).
\end{equation}
These states satisfy the orthonormality
\begin{equation}
\delta _{\tau _I \sigma _I} =\fourIdx{}{I}{}{}{\widetilde{\langle\tau _I}}|\widetilde{\sigma _I \rangle}_I =\fourIdx{(0)}{I}{}{}{\widetilde{\langle\tau _I}}|\widetilde{\sigma _I \rangle}_I^{(0)} +\fourIdx{(0)}{I}{}{}{\widetilde{\langle\tau _I}}|\widetilde{\sigma _I \rangle}_I^{(1)} +\fourIdx{(1)}{I}{}{}{\widetilde{\langle\tau _I}}|\widetilde{\sigma _I \rangle}_I^{(0)} +\mathcal{O} (\delta ^2),
\end{equation}
which yields the constraint at each order:
\begin{equation}
\fourIdx{(0)}{I}{}{}{\widetilde{\langle\tau _I}}|\widetilde{\sigma _I \rangle}_I^{(0)} =\delta _{\tau _I \sigma _I} ,\quad\fourIdx{(0)}{I}{}{}{\widetilde{\langle\tau _I}}|\widetilde{\sigma _I \rangle}_I^{(1)} +\fourIdx{(1)}{I}{}{}{\widetilde{\langle\tau _I}}|\widetilde{\sigma _I \rangle}_I^{(0)} =0.
\end{equation}
The block-product state
\begin{equation}
\widetilde{\ket{\sigma}} =\bigotimes _{I=1}^{\widetilde{L}} \widetilde{\ket{\sigma _I}}_I =\bigotimes _{I=1}^{\widetilde{L}} \left(\widetilde{\ket{\sigma _I}}_I^{(0)} +\widetilde{\ket{\sigma _I}}_I^{(1)} +\mathcal{O} (\delta ^2)\right)
\end{equation}
has the zeroth- and first-order parts
\bes
\begin{align}
\widetilde{\ket{\sigma}}^{(0)} & =\widetilde{\ket{\sigma _1}}_1^{(0)} \dotsm\widetilde{\ket{\sigma _{\widetilde{L}}}}_{\widetilde{L}}^{(0)} , \\
\widetilde{\ket{\sigma}}^{(1)} & =\sum _{I=1}^{\widetilde{L}} \widetilde{\ket{\sigma _1}}_1^{(0)} \dotsm\widetilde{\ket{\sigma _{I-1}}}_{I-1}^{(0)} \widetilde{\ket{\sigma _I}}_I^{(1)} \widetilde{\ket{\sigma _{I+1}}}_{I+1}^{(0)} \dotsm\widetilde{\ket{\sigma _{\widetilde{L}}}}_{\widetilde{L}}^{(0)} .
\end{align}
\ees

\subsection{Variational ansatz for the projector} \label{ssec.rgeqslargek_projector}

The renormalized Hamiltonian is given by $\hat{\widetilde{h}} =\hat{Q} \hat{h} \hat{Q}$. Now we need to construct the projector $\hat{Q}$. In the standard real-space RG method, the Hilbert space would be projected onto a low-energy space of the intrablock Hamiltonian after separating the Hamiltonian into intrablock and interblock Hamiltonians~\cite{Nishimori2011}. However, we do not adopt this method, because the interblock interactions are strong and diagonalization of the intrablock Hamiltonian will not yield a low-energy space of the entire system correctly (recall that the interblock operators are of $\mathcal{O} (\delta ^0)$). In order to find a low-energy space not of an intrablock Hamiltonian but of the whole Hamiltonian $\hat{h}$, we propose a new RG procedure in which the projector $\hat{Q}$ is variationally determined.

Let us construct the zeroth-order projector $\hat{Q}^{(0)} =\sum _{\sigma\in H_{\widetilde{L}}} \widetilde{\ket{\sigma}}^{(0)} \,\,\fourIdx{(0)}{}{}{}{\widetilde{\bra{\sigma}}}$. It is sufficient to take into account only the penalty Hamiltonian $\hat{h}^{(0)}$ to determine the forms of $\widetilde{\ket{\sigma _I}}_I^{(0)}$. The construction proceeds as follows:
\begin{enumerate}
\item{Each $\widetilde{\ket{\sigma _I}}_I^{(0)}$ is a superposition of $\ket{\tau _{3I-2} \tau _{3I-1} \tau _{3I}}_I$ with $(\tau _{3I-2} ,\tau _{3I-1} ,\tau _{3I})\in D'_3$ because configurations with $(\tau _{3I-2} ,\tau _{3I-1} ,\tau _{3I})\in H_3 \setminus D'_3$ pay energy penalties.}
\item{Each $\widetilde{\ket{\sigma _I}}_I^{(0)}$ does not include both components of $\Ket{\stack{\cdots\\ \uar\uar\uar}}_I$ and $\Ket{\stack{\cdots\\ \dar\dar\dar}}_I$. If $\widetilde{\ket{\sigma _I}}_I^{(0)}$ has both components of $\Ket{\stack{\cdots\\ \uar\uar\uar}}_I$ and $\Ket{\stack{\cdots\\ \dar\dar\dar}}_I$ for at least one of $\sigma _I \in H_1$, any state of length $L$ in which the state in the $I$th block is $\widetilde{\ket{\sigma _I}}_I^{(0)}$ has a component including $\dar\uar$ or $\uar\dar$ on the bottom row whether the $(3I+1)$th spin on the bottom row is up or down. This state pays a penalty.}
\item{We suppose that $\widetilde{\Ket{\stack{\uar\\ \uar}}}_I^{(0)}$ and $\widetilde{\Ket{\stack{\dar\\ \uar}}}_I^{(0)}$ are superpositions of $\Ket{\stack{\uar\dar\uar\\ \uar\uar\uar}}_I$, $\Ket{\stack{\uar\uar\uar\\ \uar\uar\uar}}_I$, $\Ket{\stack{\dar\uar\dar\\ \uar\uar\uar}}_I$, $\Ket{\stack{\uar\uar\dar\\ \uar\uar\uar}}_I$, and $\Ket{\stack{\dar\uar\uar\\ \uar\uar\uar}}_I$, while $\widetilde{\Ket{\stack{\uar\\ \dar}}}_I^{(0)} =\Ket{\stack{\uar\dar\uar\\ \dar\dar\dar}}_I$ and $\widetilde{\Ket{\stack{\dar\\ \dar}}}_I^{(0)} =\Ket{\stack{\dar\uar\dar\\ \dar\dar\dar}}_I$. The reason is that we should keep both columnar and staggered configurations to study the phase transition in the dimer limit. Note that we do not need to take superpositions of $\Ket{\stack{\uar\dar\uar\\ \dar\dar\dar}}_I$ and $\Ket{\stack{\dar\uar\dar\\ \dar\dar\dar}}_I$ because the projector $\hat{Q}^{(0)}$ is independent of what superpositions of these states are taken.}
\item{We assume that $\widetilde{\Ket{\stack{\uar\\ \uar}}}_I^{(0)}$ is a superposition of $\Ket{\stack{\uar\dar\uar\\ \uar\uar\uar}}_I$ and $\Ket{\stack{\uar\uar\uar\\ \uar\uar\uar}}_I$, while $\widetilde{\Ket{\stack{\dar\\ \uar}}}_I^{(0)}$ is a superposition of $\Ket{\stack{\uar\dar\uar\\ \uar\uar\uar}}_I$, $\Ket{\stack{\uar\uar\uar\\ \uar\uar\uar}}_I$, $\Ket{\stack{\dar\uar\dar\\ \uar\uar\uar}}_I$, $\Ket{\stack{\uar\uar\dar\\ \uar\uar\uar}}_I$, and $\Ket{\stack{\dar\uar\uar\\ \uar\uar\uar}}_I$. This assumption is motivated by the expectation that the dimer structure will be kept around the fixed point that dominates the phase transition as the RG transformation is applied many times (we need to explore a neighborhood of a fixed point to derive critical properties). The dimer structure means that the set of low-energy states has a one-to-one correspondence with the set of dimer coverings on a two-leg ladder. We can preserve the dimer structure by constructing a zeroth-order renormalized Hamiltonian $\hat{\widetilde{h}}^{(0)}$ that imposes a penalty on $\widetilde{\ket{\sigma _I \sigma _{I+1}}}_{I,I+1}^{(0)} =\widetilde{\ket{\sigma _I}}_I^{(0)} \widetilde{\ket{\sigma _{I+1}}}_{I+1}^{(0)}$ for $(\sigma _I ,\sigma _{I+1})\in H_2 \setminus D'_2$ and no penalty for $(\sigma _I ,\sigma _{I+1})\in D'_2$. To prevent $\widetilde{\Ket{\stack{\uar\uar\\ \uar\uar}}}_{I,I+1}^{(0)}$ from paying a penalty, $\widetilde{\Ket{\stack{\uar\\ \uar}}}_I^{(0)}$ does not include $\Ket{\stack{\dar\uar\dar\\ \uar\uar\uar}}_I$ or both of $\Ket{\stack{\uar\uar\dar\\ \uar\uar\uar}}_I$ and $\Ket{\stack{\dar\uar\uar\\ \uar\uar\uar}}_I$. If $\widetilde{\Ket{\stack{\uar\\ \uar}}}_I^{(0)}$ includes $\Ket{\stack{\uar\uar\dar\\ \uar\uar\uar}}_I$ $\left(\text{but neither $\Ket{\stack{\dar\uar\dar\\ \uar\uar\uar}}_I$ nor $\Ket{\stack{\dar\uar\uar\\ \uar\uar\uar}}_I$} \right)$, then $\widetilde{\Ket{\stack{\dar\\ \uar}}}_I^{(0)}$ is also a superposition of $\Ket{\stack{\uar\dar\uar\\ \uar\uar\uar}}_I$, $\Ket{\stack{\uar\uar\uar\\ \uar\uar\uar}}_I$, and $\Ket{\stack{\uar\uar\dar\\ \uar\uar\uar}}_I$ to prevent $\widetilde{\Ket{\stack{\uar\dar\\ \uar\uar}}}_{I,I+1}^{(0)}$ from paying a penalty. In this case, however, $\widetilde{\Ket{\stack{\dar\dar\\ \uar\uar}}}_{I,I+1}^{(0)}$ does not cost energy of $\mathcal{O} (\delta ^0)$ and we cannot keep the dimer structure. In addition, there is no reason why the $(3I-2)$th spin on the top row is up for any $I$. It is thus necessary to require that $\widetilde{\Ket{\stack{\uar\\ \uar}}}_I^{(0)}$ does not include $\Ket{\stack{\uar\uar\dar\\ \uar\uar\uar}}_I$. Likewise, $\widetilde{\Ket{\stack{\uar\\ \uar}}}_I^{(0)}$ will not include $\Ket{\stack{\dar\uar\uar\\ \uar\uar\uar}}_I$. Therefore, $\widetilde{\Ket{\stack{\uar\\ \uar}}}_I^{(0)}$ contains only the components $\Ket{\stack{\uar\dar\uar\\ \uar\uar\uar}}_I$ and $\Ket{\stack{\uar\uar\uar\\ \uar\uar\uar}}_I$, while $\widetilde{\Ket{\stack{\dar\\ \uar}}}_I^{(0)}$ can contain all the elements in $D'^{\,\text{columnar}}_3$. Then, $\widetilde{\Ket{\stack{\dar\dar\\ \uar\uar}}}_{I,I+1}^{(0)}$ pays a penalty and $\widetilde{\Ket{\stack{\uar\uar\\ \uar\uar}}}_{I,I+1}^{(0)}$, $\widetilde{\Ket{\stack{\dar\uar\\ \uar\uar}}}_{I,I+1}^{(0)}$, and $\widetilde{\Ket{\stack{\uar\dar\\ \uar\uar}}}_{I,I+1}^{(0)}$ do not. This means that we reproduce the dimer structure after projecting the Hilbert space.}
\item{Since $\widetilde{\ket{\sigma _I}}_I^{(0)}$ are orthonormal, we obtain
\bes
\label{eq.rgeqslargek_tildeketblock0}
\begin{align}
\widetilde{\Ket{\stack{\uar\\ \uar}}}_I^{(0)} & =\alpha _1 \Ket{\stack{\uar\dar\uar\\ \uar\uar\uar}}_I +\alpha _2 \Ket{\stack{\uar\uar\uar\\ \uar\uar\uar}}_I , \label{eq.rgeqslargek_tildeketblock0uu} \\
\widetilde{\Ket{\stack{\dar\\ \uar}}}_I^{(0)} & =-z^* \alpha _2^* \Ket{\stack{\uar\dar\uar\\ \uar\uar\uar}}_I +z^* \alpha _1^* \Ket{\stack{\uar\uar\uar\\ \uar\uar\uar}}_I +\beta _1^* \Ket{\stack{\dar\uar\dar\\ \uar\uar\uar}}_I +\beta _2^* \Ket{\stack{\uar\uar\dar\\ \uar\uar\uar}}_I +\beta _3^* \Ket{\stack{\dar\uar\uar\\ \uar\uar\uar}}_I , \label{eq.rgeqslargek_tildeketblock0du} \\
\widetilde{\Ket{\stack{\uar\\ \dar}}}_I^{(0)} & =\Ket{\stack{\uar\dar\uar\\ \dar\dar\dar}}_I , \label{eq.rgeqslargek_tildeketblock0ud} \\
\widetilde{\Ket{\stack{\dar\\ \dar}}}_I^{(0)} & =\Ket{\stack{\dar\uar\dar\\ \dar\dar\dar}}_I , \label{eq.rgeqslargek_tildeketblock0dd}
\end{align}
\ees
where the parameters $\alpha _1 ,\alpha _2 ,z,\beta _1 ,\beta _2 ,\beta _3 \in\mathbb{C}$ satisfy the normalization conditions:
\begin{align}
\fourIdx{(0)}{I}{}{}{\widetilde{\langle\sigma _I}}|\widetilde{\sigma _I \rangle}_I^{(0)} =1 \quad (\forall\sigma _I \in H_1) & \iff\left\{
\begin{aligned}
& \lvert\alpha _1 \rvert ^2 +\lvert\alpha _2 \rvert ^2 =1, \\
& \lvert z\rvert ^2 (\lvert\alpha _1 \rvert ^2 +\lvert\alpha _2 \rvert ^2)+\lvert\beta _1 \rvert ^2 +\lvert\beta _2 \rvert ^2 +\lvert\beta _3 \rvert ^2 =1
\end{aligned}
\right. \notag \\
& \iff\left\{
\begin{aligned}
& \lvert\alpha _1 \rvert ^2 +\lvert\alpha _2 \rvert ^2 =1, \\
& \lvert z\rvert ^2 +\lvert\beta _1 \rvert ^2 +\lvert\beta _2 \rvert ^2 +\lvert\beta _3 \rvert ^2 =1.
\end{aligned}
\right. \label{eq.rgeqslargek_normtildekets}
\end{align}
Equation~\eqref{eq.rgeqslargek_tildeketblock0} can be regarded as a variational ansatz for the projector $\hat{Q}^{(0)}$. We will determine the variational parameters $\alpha _1$, $\alpha _2$, $z$, $\beta _1$, $\beta _2$, and $\beta _3$ in Appendix~\ref{ssec.rgeqslargek_detvarprm}.}
\end{enumerate}

\subsection{Projection of the Hamiltonian} \label{ssec.rgeqslargek_projh}

Let us compute the renormalized Hamiltonian $\hat{\widetilde{h}} =\hat{Q} \hat{h} \hat{Q}$ up to first order. First, the zeroth-order Hamiltonian $\hat{h}^{(0)}$ is projected onto
\begin{align}
\hat{Q} \hat{h}^{(0)} \hat{Q} & =\sum _{\sigma ,\tau\in H_{\widetilde{L}}} \widetilde{\ket{\tau}} \widetilde{\bra{\tau}} \hat{h}^{(0)} \widetilde{\ket{\sigma}} \widetilde{\bra{\sigma}} \notag \\
& =\sum _{\sigma ,\tau\in H_{\widetilde{L}}} \left[\widetilde{\ket{\tau}} \left(\fourIdx{(0)}{}{}{}{\widetilde{\bra{\tau}}} \hat{h}^{(0)} \widetilde{\ket{\sigma}}^{(0)} +\fourIdx{(0)}{}{}{}{\widetilde{\bra{\tau}}} \hat{h}^{(0)} \widetilde{\ket{\sigma}}^{(1)} +\fourIdx{(1)}{}{}{}{\widetilde{\bra{\tau}}} \hat{h}^{(0)} \widetilde{\ket{\sigma}}^{(0)} \right)\widetilde{\bra{\sigma}} \right] +\mathcal{O} (\delta ^2).
\end{align}
It is convenient to write the zeroth-order Hamiltonian as a summation over the block index $I$:
\bes
\begin{align}
\hat{h}^{(0)} & =\sum _{I=1}^{\widetilde{L}} \hat{h}_{I,I+1}^{(0)} , \\
\hat{h}_{I,I+1}^{(0)} & =\sum _{(\sigma _1 ,\sigma _2)\in H_2 \setminus D'_2} k(\sigma _1 ,\sigma _2)\biggl[\frac{1}{2} (\ket{\sigma _1 \sigma _2} \bra{\sigma _1 \sigma _2})_{3I-2,3I-1} +\frac{1}{2} (\ket{\sigma _1 \sigma _2} \bra{\sigma _1 \sigma _2})_{3I-1,3I} \notag \\
& \hphantom{{} =\sum _{(\sigma _1 ,\sigma _2)\in H_2 \setminus D'_2} k(\sigma _1 ,\sigma _2)\biggl[} +(\ket{\sigma _1 \sigma _2} \bra{\sigma _1 \sigma _2})_{3I,3I+1} \notag \\
& \hphantom{{} =\sum _{(\sigma _1 ,\sigma _2)\in H_2 \setminus D'_2} k(\sigma _1 ,\sigma _2)\biggl[} +\frac{1}{2} (\ket{\sigma _1 \sigma _2} \bra{\sigma _1 \sigma _2})_{3I+1,3I+2} +\frac{1}{2} (\ket{\sigma _1 \sigma _2} \bra{\sigma _1 \sigma _2})_{3I+2,3I+3} \biggr] .
\end{align}
\ees
Since
\begin{equation}
(\ket{\rho _1 \rho _2} \bra{\rho _1 \rho _2})_{3I-2,3I-1} \widetilde{\ket{\sigma_I}}_I^{(0)} =(\ket{\rho _1 \rho _2} \bra{\rho _1 \rho _2})_{3I-1,3I} \widetilde{\ket{\sigma_I}}_I^{(0)} =0
\end{equation}
for $(\rho _1 ,\rho _2)\in H_2 \setminus D'_2$ and $\sigma _I \in H_1$, we obtain
\begin{equation}
\fourIdx{(0)}{I,I+1}{}{}{\widetilde{\bra{\tau _I \tau _{I+1}}}} \hat{h}_{I,I+1}^{(0)} \widetilde{\ket{\sigma _I \sigma _{I+1}}}_{I,I+1}^{(0)} =\sum _{(\rho _1 ,\rho _2)\in H_2 \setminus D'_2} k(\rho _1 ,\rho _2)\,\,\fourIdx{(0)}{I,I+1}{}{}{\widetilde{\bra{\tau _I \tau _{I+1}}}} (\ket{\rho _1 \rho _2} \bra{\rho _1 \rho _2})_{3I,3I+1} \widetilde{\ket{\sigma _I \sigma _{I+1}}}_{I,I+1}^{(0)}
\end{equation}
for $(\sigma _I ,\sigma _{I+1}),(\tau _I ,\tau _{I+1})\in H_2$. After calculations, we find that the matrix $\fourIdx{(0)}{I,I+1}{}{}{\widetilde{\bra{\tau _I \tau _{I+1}}}} \hat{h}_{I,I+1}^{(0)} \widetilde{\ket{\sigma _I \sigma _{I+1}}}_{I,I+1}^{(0)}$ is diagonal:
\begin{equation}
\fourIdx{(0)}{I,I+1}{}{}{\widetilde{\bra{\tau _I \tau _{I+1}}}} \hat{h}_{I,I+1}^{(0)} \widetilde{\ket{\sigma _I \sigma _{I+1}}}_{I,I+1}^{(0)} =\delta _{(\tau _I ,\tau _{I+1}),(\sigma _I ,\sigma _{I+1})} \,\,\fourIdx{(0)}{I,I+1}{}{}{\widetilde{\bra{\sigma _I \sigma _{I+1}}}} \hat{h}_{I,I+1}^{(0)} \widetilde{\ket{\sigma _I \sigma _{I+1}}}_{I,I+1}^{(0)} ,
\end{equation}
whose diagonal elements are written as
\begin{equation}
\fourIdx{(0)}{I,I+1}{}{}{\widetilde{\bra{\sigma _I \sigma _{I+1}}}} \hat{h}_{I,I+1}^{(0)} \widetilde{\ket{\sigma _I \sigma _{I+1}}}_{I,I+1}^{(0)} =\left\{
\begin{alignedat}{2}
& 0, && \quad (\sigma _I ,\sigma _{I+1})\in D'_2 , \\
& \widetilde{k} (\sigma _I ,\sigma _{I+1}), && \quad (\sigma _I ,\sigma _{I+1})\in H_2 \setminus D'_2 .
\end{alignedat}
\right.
\end{equation}
Here, $\widetilde{k} (\sigma _I ,\sigma _{I+1})$ are given by
\bes
\label{eq.rgeqslargek_dduuh0block}
\begin{align}
\widetilde{k} \left(\stack{\dar\dar\\ \uar\uar} \right) & =k\left(\stack{\dar\dar\\ \uar\uar} \right) (\lvert\beta _1 \rvert ^2 +\lvert\beta _2 \rvert ^2)(\lvert\beta _1 \rvert ^2 +\lvert\beta _3 \rvert ^2), \label{eq.rgeqslargek_dduuh0blockdduu} \\
\widetilde{k} \left(\stack{\uar\uar\\ \dar\dar} \right) & =k\left(\stack{\uar\uar\\ \dar\dar} \right) , \\
\widetilde{k} \left(\stack{\dar\dar\\ \dar\dar} \right) & =k\left(\stack{\dar\dar\\ \dar\dar} \right) , \\
\widetilde{k} \left(\stack{\uar\uar\\ \uar\dar} \right) & =k\left(\stack{\uar\uar\\ \uar\dar} \right) , \\
\widetilde{k} \left(\stack{\uar\uar\\ \dar\uar} \right) & =k\left(\stack{\uar\uar\\ \dar\uar} \right) , \\
\widetilde{k} \left(\stack{\uar\dar\\ \uar\dar} \right) & =k\left(\stack{\uar\dar\\ \uar\dar}\right) , \\
\widetilde{k} \left(\stack{\dar\uar\\ \dar\uar} \right) & =k\left(\stack{\dar\uar\\ \dar\uar}\right) , \\
\widetilde{k} \left(\stack{\dar\uar\\ \uar\dar} \right) & =k\left(\stack{\uar\uar\\ \uar\dar} \right) (\lvert z\rvert ^2 +\lvert\beta _3 \rvert ^2)+k\left(\stack{\dar\uar\\ \uar\dar} \right) (\lvert\beta _1 \rvert ^2 +\lvert\beta _2 \rvert ^2), \\
\widetilde{k} \left(\stack{\uar\dar\\ \dar\uar} \right) & =k\left(\stack{\uar\uar\\ \dar\uar} \right) (\lvert z\rvert ^2 +\lvert\beta _2 \rvert ^2)+k\left(\stack{\uar\dar\\ \dar\uar} \right) (\lvert\beta _1 \rvert ^2 +\lvert\beta _3 \rvert ^2), \\
\widetilde{k} \left(\stack{\dar\dar\\ \uar\dar} \right) & =k\left(\stack{\uar\dar\\ \uar\dar} \right) (\lvert z\rvert ^2 +\lvert\beta _3 \rvert ^2)+k\left(\stack{\dar\dar\\ \uar\dar} \right) (\lvert\beta _1 \rvert ^2 +\lvert\beta _2 \rvert ^2), \\
\widetilde{k} \left(\stack{\dar\dar\\ \dar\uar} \right) & =k\left(\stack{\dar\uar\\ \dar\uar} \right) (\lvert z\rvert ^2 +\lvert\beta _2 \rvert ^2)+k\left(\stack{\dar\dar\\ \dar\uar} \right) (\lvert\beta _1 \rvert ^2 +\lvert\beta _3 \rvert ^2). \label{eq.rgeqslargek_ddduh0blockdddu}
\end{align}
\ees
Then, we obtain
\begin{align}
\hat{Q} \hat{h}^{(0)} \hat{Q} & =\sum _{I=1}^{\widetilde{L}} \sum _{\sigma _1 ,\dots ,\sigma _{I-1} ,\sigma _{I+2} ,\dots ,\sigma _{\widetilde{L}} \in H_1} \sum _{(\sigma _I ,\sigma _{I+1})\in H_2 \setminus D'_2} \widetilde{k} (\sigma _I ,\sigma _{I+1})\widetilde{\ket{\sigma _1 \dotsm\sigma _{\widetilde{L}}}} \widetilde{\bra{\sigma _1 \dotsm\sigma _{\widetilde{L}}}} \notag \\
& \hphantom{{} = {}} +\left(\sum _{\tau\in H_{\widetilde{L}}} \sum _{\sigma\in H_{\widetilde{L}} \setminus D_{\widetilde{L}}} \widetilde{\ket{\tau}} \,\,\fourIdx{(1)}{}{}{}{\widetilde{\bra{\tau}}} \hat{h}^{(0)} \widetilde{\ket{\sigma}}^{(0)} \,\,\widetilde{\bra{\sigma}} +\mathrm{h.c.} \right) +\mathcal{O} (\delta ^2), \label{eq.rgeqslargek_Qh0Q}
\end{align}
where $\mathrm{h.c.}$ stands for the Hermitian conjugate. 
Here, we used $\sigma\in D_{\widetilde{L}} \Longrightarrow\hat{h}^{(0)} \widetilde{\ket{\sigma}}^{(0)} =0$, which follows from $\sigma\in D_{\widetilde{L}} \iff (\sigma _I ,\sigma _{I+1})\in D'_2$ ($\forall I=1,\dots ,\widetilde{L}$) and $(\sigma _I ,\sigma _{I+1})\in D'_2 \Longrightarrow\hat{h}_{I,I+1}^{(0)} \widetilde{\ket{\sigma _I \sigma _{I+1}}}_{I,I+1}^{(0)} =0$. Next, the projection of the first-order Hamiltonian $\hat{h}^{(1)}$ is given by
\begin{equation}
\hat{Q} \hat{h}^{(1)} \hat{Q} =\sum _{\sigma ,\tau\in H_{\widetilde{L}}} \widetilde{\ket{\tau}} \widetilde{\bra{\tau}} \hat{h}^{(1)} \widetilde{\ket{\sigma}} \widetilde{\bra{\sigma}} =\sum _{\sigma ,\tau\in H_{\widetilde{L}}} \widetilde{\ket{\tau}} \,\,\fourIdx{(0)}{}{}{}{\widetilde{\bra{\tau}}} \hat{h}^{(1)} \widetilde{\ket{\sigma}}^{(0)} \,\,\widetilde{\bra{\sigma}} +\mathcal{O} (\delta ^2), \label{eq.rgeqslargek_Qh1Q}
\end{equation}
where we used $\widetilde{\ket{\sigma}} =\widetilde{\ket{\sigma}}^{(0)} +\mathcal{O} (\delta ^1)$.

We can regard $\hat{Q} \hat{h}^{(0)} \hat{Q}$ and $\hat{Q} \hat{h}^{(1)} \hat{Q}$ as operators on the coarse-grained Hilbert space $\mathcal{H}_{\widetilde{L}} =\Span\{\,\widetilde{\ket{\sigma}} \,\} _{\sigma\in H_{\widetilde{L}}}$. Omitting $\mathcal{O} (\delta ^2)$ terms in Eqs.~\eqref{eq.rgeqslargek_Qh0Q} and \eqref{eq.rgeqslargek_Qh1Q}, we derive the expression of the renormalized Hamiltonian $\hat{\widetilde{h}} =\hat{Q} \hat{h} \hat{Q}$ as an operator on $\mathcal{H}_{\widetilde{L}}$:
\bes
\begin{align}
\hat{\widetilde{h}} & =\hat{\widetilde{h}}^{(0)} +\hat{\widetilde{h}}^{(1)} , \\
\hat{\widetilde{h}}^{(0)} & =\sum _{I=1}^{\widetilde{L}} \sum _{(\sigma _I ,\sigma _{I+1})\in H_2 \setminus D'_2} \widetilde{k} (\sigma _I ,\sigma _{I+1})\left(\widetilde{\ket{\sigma _I \sigma _{I+1}}} \widetilde{\bra{\sigma _I \sigma _{I+1}}} \right) _{I,I+1} , \\
\hat{\widetilde{h}}^{(1)} & =\sum _{\sigma ,\tau\in H_{\widetilde{L}}} \widetilde{\ket{\tau}} \,\,\fourIdx{(0)}{}{}{}{\widetilde{\bra{\tau}}} \hat{h}^{(1)} \widetilde{\ket{\sigma}}^{(0)} \,\,\widetilde{\bra{\sigma}} +\hat{\widetilde{o}} \hat{P} (\mathcal{D}_{\widetilde{L}}^\perp )+\hat{P} (\mathcal{D}_{\widetilde{L}}^\perp )\hat{\widetilde{o}}^\dagger , \label{eq.rgeqslargek_tildeh1}
\end{align}
\ees
where $\hat{P} (\mathcal{D}_{\widetilde{L}}^\perp )$ is the projector onto the orthogonal complement of $\mathcal{D}_{\widetilde{L}}$ and
\begin{equation}
\hat{\widetilde{o}} :=\sum _{\tau\in H_{\widetilde{L}}} \sum _{\sigma\in H_{\widetilde{L}} \setminus D_{\widetilde{L}}} \widetilde{\ket{\tau}} \,\,\fourIdx{(1)}{}{}{}{\widetilde{\bra{\tau}}} \hat{h}^{(0)} \widetilde{\ket{\sigma}}^{(0)} \,\,\widetilde{\bra{\sigma}} =\mathcal{O} (\delta ^1).
\end{equation}
Note that $\hat{\widetilde{o}} =\hat{\widetilde{o}} \hat{P} (\mathcal{D}_{\widetilde{L}}^\perp )$ because $\widetilde{\ket{\sigma}} \in\mathcal{D}_{\widetilde{L}}^\perp$ for $\sigma\in H_{\widetilde{L}} \setminus D_{\widetilde{L}}$. The operator $\hat{\widetilde{o}} \hat{P} (\mathcal{D}_{\widetilde{L}}^\perp )+\hat{P} (\mathcal{D}_{\widetilde{L}}^\perp )\hat{\widetilde{o}}^\dagger$ will be irrelevant.

It follows from Eq.~\eqref{eq.rgeqslargek_dduuh0block} and $0<k(\rho _1 ,\rho _2)=\mathcal{O} (\delta ^0)$ for $(\rho _1 ,\rho _2)\in H_2 \setminus D'_2$ that if
\begin{equation}
\lvert\beta _1 \rvert ^2 +\lvert\beta _2 \rvert ^2 >0\wedge\lvert\beta _1 \rvert ^2 +\lvert\beta _3 \rvert ^2 >0\iff\beta _1 \not= 0\vee (\beta _2 \not= 0\wedge\beta _3 \not= 0),
\end{equation}
then
\begin{equation}
\forall (\sigma _I ,\sigma _{I+1})\in H_2 \setminus D'_2 ,\quad 0<\widetilde{k} (\sigma _I ,\sigma _{I+1})=\mathcal{O} (\delta ^0).
\end{equation}
This means that the zeroth-order renormalized Hamiltonian $\hat{\widetilde{h}}^{(0)}$ imposes an energy penalty $\widetilde{k} (\sigma _I ,\sigma _{I+1})$ on any $\widetilde{\ket{\sigma _I \sigma _{I+1}}}_{I,I+1}$ with $(\sigma _I ,\sigma _{I+1})\in H_2 \setminus D'_2$, which involves a penalty on every state in $\mathcal{H}_{\widetilde{L}} \setminus\mathcal{D}_{\widetilde{L}}$.

Using Eqs.~\eqref{eq.rgeqslargek_h1}, \eqref{eq.rgeqslargek_tildeketblock0}, and \eqref{eq.rgeqslargek_tildeh1}, we obtain
\begin{align}
\hat{\widetilde{h}}^{(1)} & =\sum _{I=1}^{\widetilde{L}} \Biggl[\frac{3}{2} u\Biggl(\widetilde{\Ket{\stack{\uar\\ \uar}}} \widetilde{\Bra{\stack{\uar\\ \uar}}} +\widetilde{\Ket{\stack{\dar\\ \uar}}} \widetilde{\Bra{\stack{\dar\\ \uar}}} -\widetilde{\Ket{\stack{\uar\\ \dar}}} \widetilde{\Bra{\stack{\uar\\ \dar}}} -\widetilde{\Ket{\stack{\dar\\ \dar}}} \widetilde{\Bra{\stack{\dar\\ \dar}}} \Biggr) \notag \\
& \hphantom{{} =\sum\Biggl[} -\gamma\Biggl( 2\Re (\alpha _2^* \alpha _1)\widetilde{\Ket{\stack{\uar\\ \uar}}} \widetilde{\Bra{\stack{\uar\\ \uar}}} +2\Re [(\beta _2^* +\beta _3^*)(\beta _1 +z\alpha _1)-\lvert z\rvert ^2 \alpha _2^* \alpha _1]\widetilde{\Ket{\stack{\dar\\ \uar}}} \widetilde{\Bra{\stack{\dar\\ \uar}}} \notag \\
& \hphantom{{} =\sum\Biggl[ -\gamma\Biggl(} +[(\beta _2 +\beta _3)\alpha _2 +z(\alpha _1^2 -\alpha _2^2)]\widetilde{\Ket{\stack{\dar\\ \uar}}} \widetilde{\Bra{\stack{\uar\\ \uar}}} +[(\beta _2 +\beta _3)\alpha _2 +z(\alpha _1^2 -\alpha _2^2)]^* \widetilde{\Ket{\stack{\uar\\ \uar}}} \widetilde{\Bra{\stack{\dar\\ \uar}}} \Biggr) \notag \\
& \hphantom{{} =\sum\Biggl[} +v\Biggl( (\lvert\alpha _1 \rvert ^2 +3\lvert\alpha _2 \rvert ^2)\widetilde{\Ket{\stack{\uar\\ \uar}}} \widetilde{\Bra{\stack{\uar\\ \uar}}} +[\lvert z\rvert ^2 (3\lvert\alpha _1 \rvert ^2 +\lvert\alpha _2 \rvert ^2)-\lvert\beta _1 \rvert ^2 +\lvert\beta _2 \rvert ^2 +\lvert\beta _3 \rvert ^2]\widetilde{\Ket{\stack{\dar\\ \uar}}} \widetilde{\Bra{\stack{\dar\\ \uar}}} \notag \\
& \hphantom{{} =\sum\Biggl[ +v\Biggl(} +2z\alpha _1 \alpha _2 \widetilde{\Ket{\stack{\dar\\ \uar}}} \widetilde{\Bra{\stack{\uar\\ \uar}}} +(2z\alpha _1 \alpha _2)^* \widetilde{\Ket{\stack{\uar\\ \uar}}} \widetilde{\Bra{\stack{\dar\\ \uar}}} \Biggr) \notag \\
& \hphantom{{} =\sum\Biggl[} -\xi\Biggl( (\beta _2 +\beta _3)\alpha _1 \widetilde{\Ket{\stack{\dar\\ \uar}}} \widetilde{\Bra{\stack{\uar\\ \uar}}} +(\beta _2^* +\beta _3^*)\alpha _1^* \widetilde{\Ket{\stack{\uar\\ \uar}}} \widetilde{\Bra{\stack{\dar\\ \uar}}} -2\Re [(\beta _2^* +\beta _3^*)z\alpha _2]\widetilde{\Ket{\stack{\dar\\ \uar}}} \widetilde{\Bra{\stack{\dar\\ \uar}}} \Biggr)\Biggr] _I \notag \\
& \hphantom{{} = {}} -\sum _{I=1}^{\widetilde{L}} \xi\Biggl(\lvert\alpha _2 \rvert ^2 \beta _2^* \beta _3 \widetilde{\Ket{\stack{\uar\dar\\ \uar\uar}}} \widetilde{\Bra{\stack{\dar\uar\\ \uar\uar}}} +\lvert\alpha _2 \rvert ^2 \beta _2 \beta _3^* \widetilde{\Ket{\stack{\dar\uar\\ \uar\uar}}} \widetilde{\Bra{\stack{\uar\dar\\ \uar\uar}}} \Biggr) _{I,I+1} \notag \\
& \hphantom{{} = {}} +\hat{\widetilde{o}} \hat{P} (\mathcal{D}_{\widetilde{L}}^\perp )+\hat{P} (\mathcal{D}_{\widetilde{L}}^\perp )\hat{\widetilde{o}}^\dagger . \label{eq.rgeqslargek_tildeh1spec}
\end{align}

\subsection{Determination of the variational parameters} \label{ssec.rgeqslargek_detvarprm}

Now we determine the parameters $\alpha _1 ,\alpha _2 ,z,\beta _1 ,\beta _2 ,\beta _3 \in\mathbb{C}$. We should construct the projector $\hat{Q}$ such that the subspace $\mathcal{H}_{\widetilde{L}} =\Span\{\,\widetilde{\ket{\sigma}} \,\} _{\sigma\in H_{\widetilde{L}}} \subset\mathcal{H}_L$ is a low-energy space. One may argue that we should minimize the sum of the eigenenergies of the renormalized Hamiltonian, i.e., the trace of $\hat{\widetilde{h}}$. However, minimizing the ordinary trace $\Tr\hat{\widetilde{h}}$ is inappropriate for our purpose. We wish to study the transition between the states without energy penalties of $\mathcal{O} (\delta ^0)$. This implies that we should ignore states that cost energies of $\mathcal{O} (\delta ^0)$. Hence, we minimize the partial trace $\Tr _{\lvert\mathcal{D}_{\widetilde{L}}} \hat{\widetilde{h}}$, where $\Tr _{\lvert\mathcal{D}_{\widetilde{L}}}$ is the trace in the subspace $\mathcal{D}_{\widetilde{L}} =\Span\{\,\widetilde{\ket{\sigma}} \,\} _{\sigma\in D_{\widetilde{L}}} \subset\mathcal{H}_{\widetilde{L}}$:
\begin{equation}
\Tr _{\lvert\mathcal{D}_{\widetilde{L}}} \cdots :=\Tr\hat{P} (\mathcal{D}_{\widetilde{L}})\cdots\hat{P} (\mathcal{D}_{\widetilde{L}})=\Tr\hat{P} (\mathcal{D}_{\widetilde{L}})\cdots .
\end{equation}
We denoted the projector onto $\mathcal{D}_{\widetilde{L}}$ by $\hat{P} (\mathcal{D}_{\widetilde{L}})$. The projector $\hat{P} (\mathcal{D}_{\widetilde{L}})$ has the spectral decomposition
\begin{equation}
\hat{P} (\mathcal{D}_{\widetilde{L}})=\sum _{\sigma\in D_{\widetilde{L}}} \widetilde{\ket{\sigma}} \widetilde{\bra{\sigma}}
\end{equation}
and the partial trace can be written as
\begin{equation}
\Tr _{\lvert\mathcal{D}_{\widetilde{L}}} \cdots =\sum _{\sigma\in D_{\widetilde{L}}} \widetilde{\bra{\sigma}} \cdots\widetilde{\ket{\sigma}} .
\end{equation}

Since $\widetilde{\bra{\sigma}} \hat{\widetilde{h}}^{(0)} \widetilde{\ket{\sigma}} =0$ for any $\sigma\in D_{\widetilde{L}}$, we have $\Tr _{\lvert\mathcal{D}_{\widetilde{L}}} \hat{\widetilde{h}} =\Tr _{\lvert\mathcal{D}_{\widetilde{L}}} \hat{\widetilde{h}}^{(1)}$. To calculate the partial trace of $\hat{\widetilde{h}}^{(1)}$, we use the formulae
\bes
\begin{align}
\Tr _{\lvert\mathcal{D}_{\widetilde{L}}} \left(\widetilde{\Ket{\stack{\uar\\ \uar}}} \widetilde{\Bra{\stack{\uar\\ \uar}}} \right) _I & =F_{\widetilde{L} -1} , \\
\Tr _{\lvert\mathcal{D}_{\widetilde{L}}} \left(\widetilde{\Ket{\stack{\dar\\ \uar}}} \widetilde{\Bra{\stack{\dar\\ \uar}}} \right) _I & =F_{\widetilde{L} -3} , \\
\Tr _{\lvert\mathcal{D}_{\widetilde{L}}} \left(\widetilde{\Ket{\stack{\uar\\ \dar}}} \widetilde{\Bra{\stack{\uar\\ \dar}}} \right) _I & =\Tr _{\lvert\mathcal{D}_{\widetilde{L}}} \left(\widetilde{\Ket{\stack{\dar\\ \dar}}} \widetilde{\Bra{\stack{\dar\\ \dar}}} \right) _I =1
\end{align}
\ees
for $I=1,\dots ,\widetilde{L}$, where $F_{\widetilde{L} -1}$ and $F_{\widetilde{L} -3}$ are the Fibonacci numbers defined by Eq.~\eqref{eq.rgeqslargek_fibonacci}. For the present first-order renormalized Hamiltonian~\eqref{eq.rgeqslargek_tildeh1spec}, the partial trace of the renormalized Hamiltonian is
\begin{align}
\Tr _{\lvert\mathcal{D}_{\widetilde{L}}} \hat{\widetilde{h}} & =\widetilde{L} \biggl(\frac{3}{2} u(F_{\widetilde{L} -1} +F_{\widetilde{L} -3} -2) \notag \\
& \hphantom{{} =\widetilde{L} \biggl(} -\gamma\{ 2F_{\widetilde{L} -1} \Re (\alpha _2^* \alpha _1)+2F_{\widetilde{L} -3} \Re [(\beta _2^* +\beta _3^*)(\beta _1 +z\alpha _1)-\lvert z\rvert ^2 \alpha _2^* \alpha _1]\} \notag \\
& \hphantom{{} =\widetilde{L} \biggl(} +v\{ F_{\widetilde{L} -1} (\lvert\alpha _1 \rvert ^2 +3\lvert\alpha _2 \rvert ^2)+F_{\widetilde{L} -3} [\lvert z\rvert ^2 (3\lvert\alpha _1 \rvert ^2 +\lvert\alpha _2 \rvert ^2)-\lvert\beta _1 \rvert ^2 +\lvert\beta _2 \rvert ^2 +\lvert\beta _3 \rvert ^2]\} \notag \\
& \hphantom{{} =\widetilde{L} \biggl(} +2\xi F_{\widetilde{L} -3} \Re [(\beta _2^* +\beta _3^*)z\alpha _2]\biggr) .
\end{align}
Minimization of $\Tr _{\lvert\mathcal{D}_{\widetilde{L}}} \hat{\widetilde{h}}$ is equivalent to minimizing the function
\begin{align}
f_{\gamma ,v,\xi} (\alpha _1 ,\alpha _2 ,z,\beta _1 ,\beta _2 ,\beta _3) & =-\frac{1}{2} \{\gamma [(\varphi _{\widetilde{L}}^2 -\lvert z\rvert ^2)\alpha _2^* \alpha _1 +(\varphi _{\widetilde{L}}^2 -\lvert z\rvert ^2)\alpha _2 \alpha _1^* \notag \\
& \hphantom{{} =-\frac{1}{2} \{\gamma [} +(\beta _2^* +\beta _3^*)(\beta _1 +z\alpha _1)+(\beta _2 +\beta _3)(\beta _1^* +z^* \alpha _1^*)] \notag \\
& \hphantom{{} =-\frac{1}{2} \{} +2v[(\varphi _{\widetilde{L}}^2 -\lvert z\rvert ^2)\lvert\alpha _1 \rvert ^2 +\lvert\beta _1 \rvert ^2] \notag \\
& \hphantom{{} =-\frac{1}{2} \{} -\xi [(\beta _2^* +\beta _3^*)z\alpha _2 +(\beta _2 +\beta _3)z^* \alpha _2^*]\}
\end{align}
under the constraint given by Eq.~\eqref{eq.rgeqslargek_normtildekets}, where $\varphi _{\widetilde{L}}^2 :=F_{\widetilde{L} -1} /F_{\widetilde{L} -3}$. Since the difference of $\varphi _{\widetilde{L}}^2$ from the golden ratio squared $\varphi ^2$ becomes exponentially small for large system size $L$, we apply the approximation $\varphi _{\widetilde{L}}^2 \simeq\varphi ^2$. The function $f_{\gamma ,v,\xi}$ is rewritten as
\begin{align}
f_{\gamma ,v,\xi} (\alpha _1 ,\alpha _2 ,z,\beta _1 ,\beta _2 ,\beta _3) & =-\frac{1}{2} \{\gamma [(\varphi ^2 -\lvert z\rvert ^2)\alpha _2^* \alpha _1 +(\varphi ^2 -\lvert z\rvert ^2)\alpha _2 \alpha _1^* \notag \\
& \hphantom{{} =-\frac{1}{2} \{\gamma [} +(\beta _2^* +\beta _3^*)(\beta _1 +z\alpha _1)+(\beta _2 +\beta _3)(\beta _1^* +z^* \alpha _1^*)] \notag \\
& \hphantom{{} =-\frac{1}{2} \{} +2v[(\varphi ^2 -\lvert z\rvert ^2)\lvert\alpha _1 \rvert ^2 +\lvert\beta _1 \rvert ^2] \notag \\
& \hphantom{{} =-\frac{1}{2} \{} -\xi [(\beta _2^* +\beta _3^*)z\alpha _2 +(\beta _2 +\beta _3)z^* \alpha _2^*]\} .
\end{align}
Minimizing this function subject to the constraint~\eqref{eq.rgeqslargek_normtildekets} determines the variational parameters $\alpha _1$, $\alpha _2$, $z$, $\beta _1$, $\beta _2$, and $\beta _3$.

\subsection{RG equations} \label{ssec.rgeqslargek_rgeq}

We assume that the variational parameters $\alpha _1$, $\alpha _2$, $z$, $\beta _1$, $\beta _2$, and $\beta _3$ are real and that $\beta _2 =\beta _3$ for the sake of symmetry. Denote the Pauli operators in the basis $\{\,\widetilde{\ket{\sigma _{a,I}}}_{a,I} \,\} _{\sigma _{a,I} =\uar ,\dar}$ by $\hat{\widetilde{X}}_{a,I}$, $\hat{\widetilde{Y}}_{a,I}$, and $\hat{\widetilde{Z}}_{a,I}$ for $a=\rmt ,\rmb$ and $I=1,\dots ,\widetilde{L}$. It follows from Eq.~\eqref{eq.rgeqslargek_tildeh1spec} that
\begin{align}
\hat{\widetilde{h}}^{(1)} & =\sum _{I=1}^{\widetilde{L}} \Biggl(\frac{1}{2} \{ 3u-\gamma [2\beta _2 (\beta _1 +z\alpha _1)+(1-z^2)\alpha _1 \alpha _2]+v[2-(1-z^2)\alpha _1^2 -\beta _1^2]+2\xi\beta _2 z\alpha _2 \}\hat{\widetilde{Z}}_{\rmb ,I} \notag \\
& \hphantom{{} =\sum\Biggl(} +\{\gamma [2\beta _2 (\beta _1 +z\alpha _1)-(1+z^2)\alpha _1 \alpha _2]+v[1-(1+z^2)\alpha _1^2 +\beta _1^2]-2\xi\beta _2 z\alpha _2 \}\hat{\widetilde{Z}}_{\rmt ,I} \frac{1+\hat{\widetilde{Z}}_{\rmb ,I}}{2} \notag \\
& \hphantom{{} =\sum\Biggl(} -\{\gamma [2\beta _2 \alpha _2 +z(\alpha _1^2 -\alpha _2^2)]-2vz\alpha _1 \alpha _2 +2\xi\beta _2 \alpha _1 \}\hat{\widetilde{X}}_{\rmt ,I} \frac{1+\hat{\widetilde{Z}}_{\rmb ,I}}{2} \notag \\
& \hphantom{{} =\sum\Biggl(} -\xi\alpha _2^2 \beta _2^2 \left(\widetilde{\Ket{\stack{\uar\dar\\ \uar\uar}}} \widetilde{\Bra{\stack{\dar\uar\\ \uar\uar}}} +\widetilde{\Ket{\stack{\dar\uar\\ \uar\uar}}} \widetilde{\Bra{\stack{\uar\dar\\ \uar\uar}}} \right) _{I,I+1} \notag \\
& \hphantom{{} =\sum\Biggl(} +\frac{1}{2} \{ -\gamma [2\beta _2 (\beta _1 +z\alpha _1)+(1-z^2)\alpha _1 \alpha _2]+v[2-(1-z^2)\alpha _1^2 -\beta _1^2]+2\xi\beta _2 z\alpha _2 \}\Biggr) \notag \\
& \hphantom{{} = {}} +\hat{\widetilde{o}} \hat{P} (\mathcal{D}_{\widetilde{L}}^\perp )+\hat{P} (\mathcal{D}_{\widetilde{L}}^\perp )\hat{\widetilde{o}}^\dagger .
\end{align}

Omitting the term proportional to the identity in $\hat{\widetilde{h}}^{(1)}$, we arrive at the expression of the renormalized Hamiltonian:
\bes
\begin{align}
\hat{\widetilde{h}} & =\hat{\widetilde{h}}^{(0)} +\hat{\widetilde{h}}^{(1)} , \\
\hat{\widetilde{h}}^{(0)} & =\sum _{I=1}^{\widetilde{L}} \sum _{(\sigma _I ,\sigma _{I+1})\in H_2 \setminus D'_2} \widetilde{k} (\sigma _I ,\sigma _{I+1})\left(\widetilde{\ket{\sigma _I \sigma _{I+1}}} \widetilde{\bra{\sigma _I \sigma _{I+1}}} \right) _{I,I+1} , \\
\hat{\widetilde{h}}^{(1)} & =\sum _{I=1}^{\widetilde{L}} \left[\frac{\widetilde{u}}{2} \hat{\widetilde{Z}}_{\rmb ,I} -\widetilde{\gamma} \hat{\widetilde{X}}_{\rmt ,I} \frac{1+\hat{\widetilde{Z}}_{\rmb ,I}}{2} +\widetilde{v} \hat{\widetilde{Z}}_{\rmt ,I} \frac{1+\hat{\widetilde{Z}}_{\rmb ,I}}{2} -\widetilde{\xi} \left(\widetilde{\Ket{\stack{\uar\dar\\ \uar\uar}}} \widetilde{\Bra{\stack{\dar\uar\\ \uar\uar}}} +\widetilde{\Ket{\stack{\dar\uar\\ \uar\uar}}} \widetilde{\Bra{\stack{\uar\dar\\ \uar\uar}}} \right) _{I,I+1} \right] +\hat{\widetilde{o}} \hat{P} (\mathcal{D}_{\widetilde{L}}^\perp )+\hat{P} (\mathcal{D}_{\widetilde{L}}^\perp )\hat{\widetilde{o}}^\dagger .
\end{align}
\ees
The renormalized coupling constants $\widetilde{u}$, $\widetilde{\gamma}$, $\widetilde{v}$, and $\widetilde{\xi}$ are given by the following RG equations:
\bes
\begin{align}
\widetilde{u} & =3u-\gamma [2\beta _2 (\beta _1 +z\alpha _1)+(1-z^2)\alpha _1 \alpha _2]+v[2-(1-z^2)\alpha _1^2 -\beta _1^2]+2\xi\beta _2 z\alpha _2 , \\
\widetilde{\gamma} & =\gamma [2\beta _2 \alpha _2 +z(\alpha _1^2 -\alpha _2^2) ]-2vz\alpha _1 \alpha _2 +2\xi\beta _2 \alpha _1 , \\
\widetilde{v} & =\gamma [2\beta _2 (\beta _1 +z\alpha _1)-(1+z^2)\alpha _1 \alpha _2]+v[1-(1+z^2)\alpha _1^2 +\beta _1^2]-2\xi\beta _2 z\alpha _2 , \\
\widetilde{\xi} & =\xi\alpha _2^2 \beta _2^2 ,
\end{align}
\ees
where $\alpha _1 ,\alpha _2 ,z,\beta _1 ,\beta _2 \in\mathbb{R}$ are the arguments minimizing the function
\begin{equation}
f_{\gamma ,v,\xi} (\alpha _1 ,\alpha _2 ,z,\beta _1 ,\beta _2)=-\gamma [(\varphi ^2 -z^2)\alpha _1 \alpha _2 +2\beta _2 (\beta _1 +z\alpha _1)]-v[(\varphi ^2 -z^2)\alpha _1^2 +\beta _1^2]+2\xi\beta _2 z\alpha _2
\end{equation}
under the constraint $\alpha _1^2 +\alpha _2^2 =z^2 +\beta _1^2 +2\beta _2^2 =1$. As explained in Sec.~\ref{ssec.rsrg_largek}, there are four optimal sets of the variational parameters due to the invariance of the minimized function and the constraint under the transformations $(\alpha _1 ,\alpha _2 ,z)\mapsto (-\alpha _1 ,-\alpha _2 ,-z)$ and $(z,\beta _1 ,\beta _2)\mapsto (-z,-\beta _1 ,-\beta _2)$, which correspond to multiplying the variational states $\widetilde{\Ket{\stack{\uar\\ \uar}}}_I^{(0)}$ and $\widetilde{\Ket{\stack{\dar\\ \uar}}}_I^{(0)}$ by a phase factor $-1$, respectively. The renormalized penalty constants $\widetilde{k} (\sigma _1 ,\sigma _2)$ are given by Eq.~\eqref{eq.rgeqslargek_dduuh0block}, but their specific values are not needed to analyze critical properties at leading order in $\delta$.

Now we perform the RG transformation repeatedly. Denoting the number of RG steps by $l$, we write the coupling constants as $U(l)=Ku(l)$, $\Gamma (l)=K\gamma (l)$, $V(l)=Kv(l)$, and $\Xi (l)=K\xi (l)$, and the variational parameters minimizing $f_{\gamma (l),v(l),\xi (l)} (\alpha _1 ,\alpha _2 ,z,\beta _1 ,\beta _2)$ as $\alpha _1 (l)$, $\alpha _2 (l)$, $z(l)$, $\beta _1 (l)$, and $\beta _2 (l)=\beta _3 (l)$ (one of the four optimal sets of the variational parameters is chosen). The coupling constants $U(l)$, $\Gamma (l)$, $V(l)$, and $\Xi (l)$ satisfy the recurrence relations~\eqref{eq.rsrg_rglargek}. We plot the coupling constants and the variational parameters as functions of $l$ for several sets of the bare couplings $(U(0),\Gamma (0),V(0),\Xi (0))=(\overline{U}_{\Gamma (0),\Xi (0)} ,\Gamma (0),0,\Xi (0))$ in Figs.~\ref{fig.rgeqslargek_runcplprmx0}--\ref{fig.rgeqslargek_runcplprmxp}, where $\overline{U} (l):=U(0)-3^{-l} U(l)$ satisfies the recurrence relation~\eqref{eq.rsrg_rgblargek} and $\overline{U}_{\Gamma (0),\Xi (0)} :=\lim _{l\to\infty} \overline{U} (l)$ determines the first-order transition point as indicated in Sec.~\ref{ssec.rsrg_largek}. Figure~\ref{fig.rgeqslargek_runcplprmx0} is for the case of no $XX$ interactions $\Xi (0)=0$, Fig.~\ref{fig.rgeqslargek_runcplprmxm} for the case of the antiferromagnetic $XX$ interactions on the top row $\Xi (0)<0$, and Fig.~\ref{fig.rgeqslargek_runcplprmxp} for the case of the ferromagnetic $XX$ interactions on the top row $\Xi (0)>0$. Note that $\overline{U} (l)$, $\Gamma (l)$, $V(l)$, and $\Xi (l)$ as well as all the variational parameters do not depend on $U(0)$, and thus $\overline{U}_{\Gamma (0),\Xi (0)}$ is determined only by $\Gamma (0)$ and $\Xi (0)$ [$V(0)$ is fixed to zero]. It turns out that $\Gamma (l)$ and $\Xi (l)$ vanish in the limit $l\to\infty$ while $\overline{U} (l)$ and $V(l)$ converge to finite positive values. In addition, $U(l)$ converges to zero when $U(0)=\overline{U}_{\Gamma (0),\Xi (0)}$.
\begin{figure*}
\includegraphics{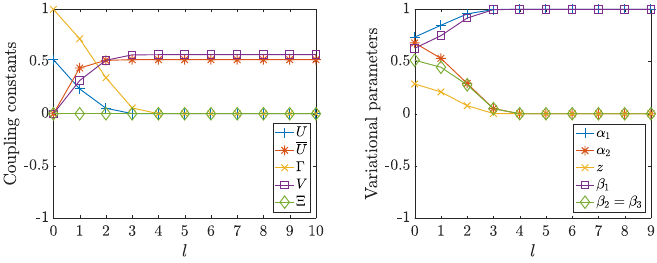}
\caption{Coupling constants $U(l)$, $\overline{U} (l)$, $\Gamma (l)$, $V(l)$, and $\Xi (l)$ and variational parameters $\alpha _1 (l)$, $\alpha _2 (l)$, $z(l)$, $\beta _1 (l)$, and $\beta _2 (l)=\beta _3 (l)$ as functions of $l$ for the bare couplings $U(0)=\overline{U}_{\Gamma (0),\Xi (0)} =0.51773501$, $\Gamma (0)=1$, $V(0)=0$, and $\Xi (0)=0$. The left graph is identical to Fig.~\ref{fig.rsrg_runcpllargek}(b) except that the present figure does not contain $U(l)$ for $U(0)\not=\overline{U}_{\Gamma (0),\Xi (0)}$.} \label{fig.rgeqslargek_runcplprmx0}
\end{figure*}
\begin{figure*}
\includegraphics{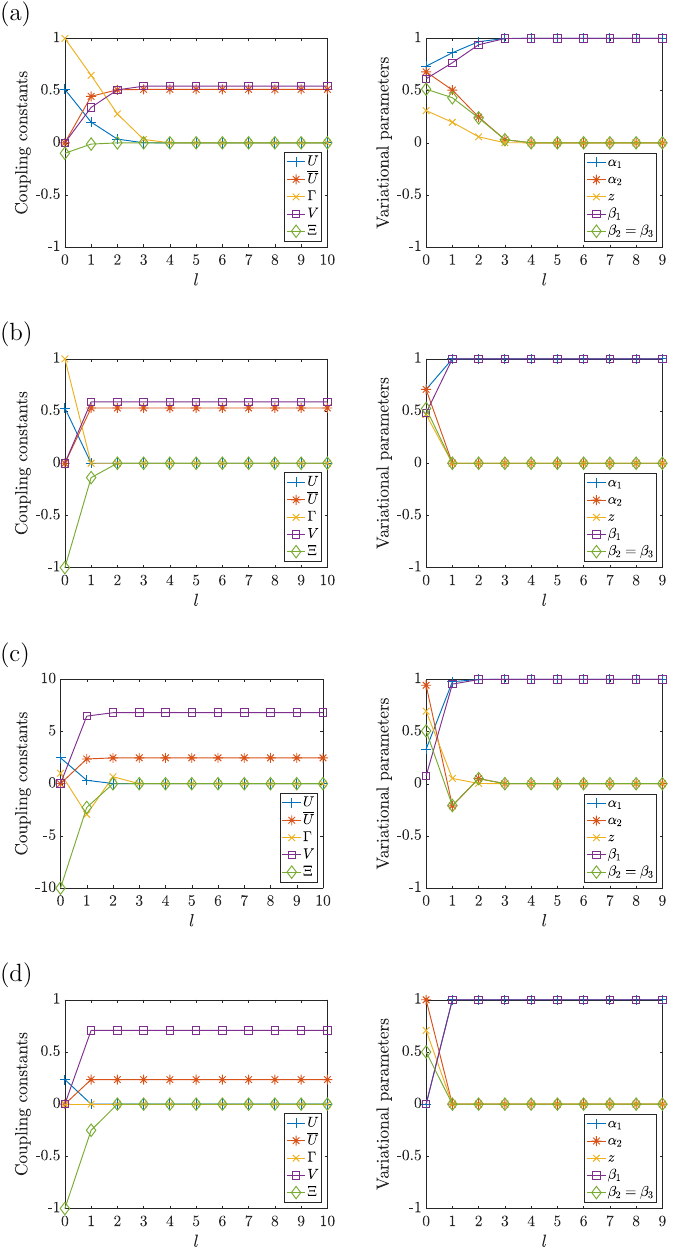}
\caption{Coupling constants $U(l)$, $\overline{U} (l)$, $\Gamma (l)$, $V(l)$, and $\Xi (l)$ and variational parameters $\alpha _1 (l)$, $\alpha _2 (l)$, $z(l)$, $\beta _1 (l)$, and $\beta _2 (l)=\beta _3 (l)$ as functions of $l$ for $\Xi (0)<0$. We set $(U(0),\Gamma (0),V(0),\Xi (0))=(0.51144374,1,0,-0.1)$ in (a), $(0.52956828,1,0,-1)$ in (b), $(2.47583092,1,0,-10)$ in (c), and $(0.23570226,0,0,-1)$ in (d), all of which satisfy $U(0)=\overline{U}_{\Gamma (0),\Xi (0)}$. The left graph of (b) is identical to Fig.~\ref{fig.rsrg_runcpllargek}(a) except that the present figure does not contain $U(l)$ for $U(0)\not=\overline{U}_{\Gamma (0),\Xi (0)}$.} \label{fig.rgeqslargek_runcplprmxm}
\end{figure*}
\begin{figure*}
\includegraphics{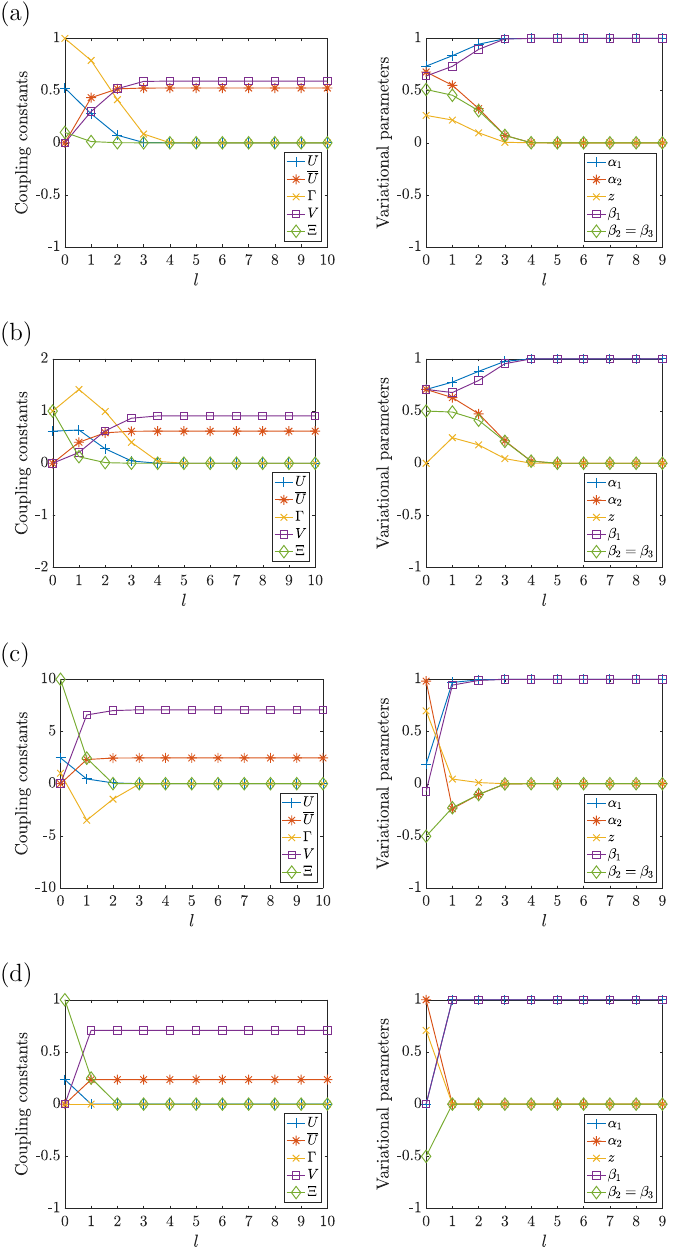}
\caption{Coupling constants $U(l)$, $\overline{U} (l)$, $\Gamma (l)$, $V(l)$, and $\Xi (l)$ and variational parameters $\alpha _1 (l)$, $\alpha _2 (l)$, $z(l)$, $\beta _1 (l)$, and $\beta _2 (l)=\beta _3 (l)$ as functions of $l$ for $\Xi (0)>0$. We set $(U(0),\Gamma (0),V(0),\Xi (0))=(0.52489031,1,0,0.1)$ in (a), $(0.61474501,1,0,1)$ in (b), $(2.46771358,1,0,10)$ in (c), and $(0.23570226,0,0,1)$ in (d), all of which satisfy $U(0)=\overline{U}_{\Gamma (0),\Xi (0)}$. The left graph of (b) is identical to Fig.~\ref{fig.rsrg_runcpllargek}(c) except that the present figure does not contain $U(l)$ for $U(0)\not=\overline{U}_{\Gamma (0),\Xi (0)}$.} \label{fig.rgeqslargek_runcplprmxp}
\end{figure*}

\end{widetext}

\section{Derivation of RG equations in the limit of small frustration} \label{sec.rgeqslargeu}

We apply the standard real-space RG method~\cite{Nishimori2011} to the frustrated Ising ladder~\eqref{eq.model_hfrustisinglad} in the limit of small frustration, or the limit of a large longitudinal field on the bottom row $U\to\infty$. The model reduces to the ferromagnetic Ising chain with a transverse field and $XX$ interactions~\eqref{eq.rsrg_hchainferrolargeu} after taking the limit $U\to\infty$ and performing the gauge transformation that changes the antiferromagnetic $ZZ$ interactions into ferromagnetic ones.

We partition the chain into $\widetilde{L} =L/2$ blocks and split the Hamiltonian into intrablock and interblock Hamiltonians:
\begin{equation}
\hat{H} =\sum _{I=1}^{\widetilde{L}} (\hat{H}_I^\text{intra} +\hat{H}_{I,I+1}^\text{inter}),
\end{equation}
where
\bes
\begin{align}
\hat{H}_I^\text{intra} & =-K\hat{Z}_{2I-1} \hat{Z}_{2I} -\Gamma\hat{X}_{2I-1} -\Xi\hat{X}_{2I-1} \hat{X}_{2I} , \\
\hat{H}_{I,I+1}^\text{inter} & =-K\hat{Z}_{2I} \hat{Z}_{2I+1} -\Gamma\hat{X}_{2I} -\Xi\hat{X}_{2I} \hat{X}_{2I+1} .
\end{align}
\ees
The intrablock Hamiltonian $\hat{H}_I^\text{intra}$, which is formed by the ($2I-1$)th and ($2I$)th spins, has the eigenvalues
\begin{equation}
\varepsilon _{s_1 s_2} =-s_1 \sqrt{K^2 +\Gamma ^2} -s_2 \Xi\quad (s_1 ,s_2 =\pm )
\end{equation}
and the corresponding eigenvectors
\bes
\begin{align}
\ket{++} & =+c_+ \ket{\rar\rar} +c_- \ket{\lar\lar} , \\
\ket{-+} & =-c_- \ket{\rar\rar} +c_+ \ket{\lar\lar} , \\
\ket{+-} & =+c_+ \ket{\rar\lar} +c_- \ket{\lar\rar} , \\
\ket{--} & =-c_- \ket{\rar\lar} +c_+ \ket{\lar\rar} .
\end{align}
\ees
Here, $\ket{\rar} :=(\ket{\uar} +\ket{\dar})/\sqrt{2}$ and $\ket{\lar} :=(\ket{\uar} -\ket{\dar})/\sqrt{2}$ are the eigenvectors of the Pauli operator $\hat{X}$ and
\begin{equation}
c_\pm :=\sqrt{\frac{1}{2} \left( 1\pm\frac{\Gamma}{\sqrt{K^2 +\Gamma ^2}} \right)} .
\end{equation}

Now we assume $\lvert\Xi\rvert <\sqrt{K^2 +\Gamma ^2}$ and project the Hilbert space onto the two-dimensional low-energy subspace of the intrablock Hamiltonian $\hat{H}_I^\text{intra}$. Since the two lowest eigenvalues of $\hat{H}_I^\text{intra}$ are $\varepsilon _{+\pm}$, the projector is given by
\begin{equation}
\hat{Q} =\bigotimes _{I=1}^{\widetilde{L}} \hat{Q}_I ,\quad\hat{Q}_I =(\ket{++} \bra{++} +\ket{+-} \bra{+-})_{2I-1,2I} .
\end{equation}
Calculating the projections of operators results in
\bes
\begin{align}
\hat{Q}_I \hat{H}_I^\text{intra} \hat{Q}_I & =\varepsilon _{++} \ket{++} \bra{++} +\varepsilon _{+-} \ket{+-} \bra{+-} , \\
\hat{Q}_I \hat{Z}_{2I-1} \hat{Q}_I & =2c_+ c_- (\ket{+-} \bra{++} +\ket{++} \bra{+-}), \\
\hat{Q}_I \hat{Z}_{2I} \hat{Q}_I & =\ket{+-} \bra{++} +\ket{++} \bra{+-} , \\
\hat{Q}_I \hat{X}_{2I-1} \hat{Q}_I & =(c_+^2 -c_-^2)(\ket{++} \bra{++} +\ket{+-} \bra{+-}) , \\
\hat{Q}_I \hat{X}_{2I} \hat{Q}_I & =(c_+^2 -c_-^2)(\ket{++} \bra{++} -\ket{+-} \bra{+-}).
\end{align}
\ees
We define $\widetilde{\ket{\uar}} :=(\ket{++} +\ket{+-})/\sqrt{2}$ and $\widetilde{\ket{\dar}} :=(\ket{++} -\ket{+-})/\sqrt{2}$ and denote the Pauli operators in the basis $\{\,\widetilde{\ket{\sigma}}_I \,\} _{\sigma =\uar ,\dar}$ by $\hat{\widetilde{X}}_I$, $\hat{\widetilde{Y}}_I$, and $\hat{\widetilde{Z}}_I$. Then, we obtain the renormalized Hamiltonian $\hat{\widetilde{H}} =\hat{Q} \hat{H} \hat{Q}$ as an operator on the coarse-grained space $\Span\{\,\bigotimes _{I=1}^{\widetilde{L}} \widetilde{\ket{\sigma _I}}_I \,\} _{\sigma _1 ,\dots ,\sigma _{\widetilde{L}} =\uar ,\dar}$:
\begin{equation}
\hat{\widetilde{H}} =\sum _{I=1}^{\widetilde{L}} (-\widetilde{K} \hat{\widetilde{Z}}_I \hat{\widetilde{Z}}_{I+1} -\widetilde{\Gamma} \hat{\widetilde{X}}_I -\widetilde{\Xi} \hat{\widetilde{X}}_I \hat{\widetilde{X}}_{I+1}),
\end{equation}
where we set
\bes
\begin{align}
\widetilde{K} & =\frac{K^2}{\sqrt{K^2 +\Gamma ^2}} , \\
\widetilde{\Gamma} & =\frac{\Gamma ^2}{\sqrt{K^2 +\Gamma ^2}} +\Xi\left( 1+\frac{\Gamma ^2}{K^2 +\Gamma ^2} \right) , \\
\widetilde{\Xi} & =0
\end{align}
\ees
and ignored a constant energy difference. Note that the above choice of the basis $\{\,\widetilde{\ket{\sigma}}_I \,\} _{\sigma =\uar ,\dar}$ makes it possible to represent the renormalized Hamiltonian as a summation of the same operators as those in the bare Hamiltonian~\eqref{eq.rsrg_hchainferrolargeu}. Finally we derive the RG equations
\begin{equation}
\widetilde{\gamma} =\gamma ^2 +\frac{\xi (1+2\gamma ^2)}{\sqrt{1+\gamma ^2}} ,\quad\widetilde{\xi} =0,
\end{equation}
where $(\gamma ,\xi )=(\Gamma /K,\Xi /K)$ and $(\widetilde{\gamma} ,\widetilde{\xi})=(\widetilde{\Gamma} /\widetilde{K} ,\widetilde{\Xi} /\widetilde{K})$.


%

\end{document}